\documentclass{elsart}
\usepackage{amssymb,psfig,graphics}

\renewcommand{\theequation}{\arabic{section}.\arabic{equation}}

\newcommand{\vek}[1]{\mbox{\boldmath$#1$}}
\newcommand{\vekind}[1]{\mbox{\scriptsize\boldmath$#1$}}
\newcommand{\psl}{p\hspace{-1.07ex}/}

\newcommand{\ksl}{\mbox{k\hspace{-1.07ex}/}}

\newcommand{\nablaleftright}{\stackrel{\leftrightarrow}{\nabla}}

\newcommand{\difleftright}{\stackrel{\leftrightarrow}{\partial}}
\newcommand{\difleft}{\stackrel{\leftarrow}{\partial}}
\newcommand{\difright}{\stackrel{\rightarrow}{\partial}}

\def\trans{\mbox{\tiny$\bot$}} 
\def\longi{\mbox{\tiny$\|$}}   

\def\spar#1#2#3#4{            
\stackrel{\hspace*{#1 pt}\vbox{\vspace*{-35pt}\hrule\hbox{\vrule
\mbox{\tiny $\!\vert$\hspace*{#2 pt}}
\vrule}}}{#3#4}}

\newcommand{\Norder}{\mbox{\boldmath$:$}}

\begin{document}

\begin{frontmatter}
\title{Covariant Linear Response Theory of Relativistic QED Plasmas}

\author[Rostock]{A. H\"oll\thanksref{Arne}}
\author[Moscow]{V.G. Morozov\thanksref{Vladimir}}
\author[Rostock]{G. R\"opke\thanksref{Gerd}}
\address[Rostock]{Physics Department, University of Rostock,
Universit\"atsplatz 3, D-18051 Rostock, Germany}
\address[Moscow]{Moscow State Institute of Radioengineering,
Electronics, and Automation, 117454 Vernadsky Prospect 78, Moscow, Russia}
\thanks[Arne]{hoell@darss.mpg.uni-rostock.de}
\thanks[Vladimir]{vmorozov@orc.ru}
\thanks[Gerd]{gerd@darss.mpg.uni-rostock.de}

\begin{abstract}
We start from the QED Lagrangian to describe a charged many-particle
system coupled to the radiation field.
A covariant density matrix approach to kinetic theory of QED plasmas,
subjected to a strong external electro-magnetic field has recently been
developed~\cite{hoell1,hoell2}.
We use the hyperplane formalism in order to perform a manifest
covariant quantization and to implement initial correlations to the
solution of the Liouville-von Neumann equation.
A perturbative expansion in orders of the fine structure constant for the
correlation functions as well as the statistical operator is applied.
The non-equilibrium state of the system is given within generalized
linear response theory.
Expressions for the susceptibility tensor, describing the plasma
response, are calculated within different approximations, like the RPA
approximation or considering collisions within the Born-approximation.
In particular, the process of relativistic inverse bremsstrahlung in a
plasma is discussed.
\end{abstract}

\end{frontmatter}

\setcounter{equation}{0}
\section{Introduction}

In recent years the theoretical study of dense relativistic plasmas is
of increasing interest. Such plasmas are not only limited to
astrophysics, but can nowadays be produced by high-intense short-pulse
lasers~\cite{Sprangle90,Gibbon1_96,Brabec1_00}.
In view of the inertial confinement fusion, one has to consider a
plasma under extreme conditions which is created by a strong external
field. This new experimental progress needs a systematic approach
based on quantum electrodynamics and methods of non-equilibrium
statistical mechanics.

Considerable  attention has been focused on a kinetic approach,
formulated for the fermionic Wigner function.
Using the Wigner operator defined in four-dimensional momentum
space~\cite{DeGroot80,CarruthersZachariasen83,VasakGyulassyElze87}, a
manifestly covariant mean-field kinetic equation can be derived from
the Heisenberg equations of motion for the field operators.
In this approach, however, it is difficult to formulate an initial value
problem for the kinetic equation since the four-dimensional Fourier
transformation of the covariant Wigner function includes integration
of two-point correlation functions over time.
Describing modern pump-and-probe laser experiments by an initial value
problem can lead to a significant simplification, since the pump and
probe process can be described separately. However, at short
time scales the plasma response depends on its initial
correlation, which have to be included appropriately.

An initial value problem can be formulated within a {\em one-time\/}
formulation, where the field operators are taken at the same time and
only the spatial Fourier transformation is performed.
In the context of QED, the one-time formulation was proposed by
Bialynicki-Birula et al.~\cite{BGR91}
and used successfully in their study of the QED vacuum.
Within this approach
the one-time Wigner function has a direct physical interpretation and
allows to calculate local observables, such as the charge density and
the current density.
The description  in terms of one-time quantities is quite natural in
kinetic theory based on the von Neumann equation for the statistical
operator and provides a consistent account of causality in collision
integrals.

It should be noted, however, that the one-time description does not
contain the complete information about one-particle dynamics. Spectral
properties of correlation functions are naturally described in terms of
two-point Green's functions which are closely related to the covariant
Wigner function.
In non-relativistic kinetic theory, where two-time correlation
functions can, in principle, be reconstructed from the one-time Wigner
function by solving integral equations which follow from the Dyson
equation for non-equilibrium Green's functions~\cite{Lipavsky86}, this
problem can be overcome.
In a relativistic theory it was suggested~\cite{Heinz} that 
an expansion in terms of energy-moments can recover the complete
spectral information within a one-time formulation.
The discussion of spectral properties will not be issued in this work.
Recently the aspect of relativistic kinetic theory was
studied within the  mean-field
approximation~\cite{Heinz,ZhuangHeinz98}.

In this work we follow the studies presented in~\cite{hoell1,hoell2}
where a density matrix
approach to kinetic theory of QED plasmas subjected to a
strong electromagnetic field was considered.
The BGR scheme~\cite{BGR91} was generalized in two aspects.
First, the one-time formalism was presented in a {\em covariant\/}
form. This removes a drawback of the BGR theory which is
not manifestly covariant.
The covariant formulation will be performed using the 
{\em hyperplane formalism}. 
Second, we demonstrate how the non-equilibrium statistical operator
can be expressed within the generalized linear response theory in a
hyperplane formalism. 
The linear response theory was successfully applied in many fields of
modern physics and can serve as tool to investigate the rather
complicated structure of general kinetic equations. Here we are
interested in the response of the plasma to an external perturbation,
caused by the laser pulse.
Considering the external field and the response of the system as a
small perturbation, the non-equilibrium statistical operator is
expanded up to linear order with respect to the external field.
It is clear, that this approximation is only reliable for moderate
laser intensities, but will break down for strong fields.
The method applied here was successfully used in
non-relativistic theory~\cite{Gerd,ZubMorRoep1}.

Having determined the non-equilibrium statistical operator, response
functions like the susceptibility tensor is given.
As an illustration, the relativistic susceptibility tensor is derived in the
mean-field approximation. 
In the present work we also give first results beyond the RPA where
interaction processes between the charged particles and the EM field
are considered. 
The susceptibility tensor, given by the current-current correlation
function, is most appropriately expressed in terms of the force-force
correlation function, which allows for a well defined perturbation
expansion with respect to the interaction. It is shown how 
the absorption coefficient is related to the imaginary part of the
force-force correlation function in lowest order perturbation theory.
In this approximation the absorption of an external field is
treated. As well known, this first
order process is possible only in combination with the Coulomb interaction
in order to obey conservation laws. In Born approximation, considered
here, the inverse bremsstrahlung is obtained. We give explicit expressions for
plasmas near the equilibrium. 
A final expression for the inverse
bremsstrahlung can be compared with other results, like relativistic
S-matrix calculations in vacuum~\cite{Heitler} or with non-relativistic
results~\cite{August}. 

The paper is organized as follows. In Section~2 we demonstrate how the
quantum Hamiltonian defined on a hyperplane can be obtained from the
classical QED Lagrangian. Major points like the gauge-fixing or the
canonical quantization on the hyperplane is reviewed. Further the
relativistic Liouville-von Neumann equation is formulated on the plane. 
In Section~3 it is shown how the Liouville-von Neumann equation can be
solved in linear response, fixing some initial distribution.
The self-consistency relation leads to a response equation, which
defines the four dimensional susceptibility tensor. The result for the
susceptibility tensor in RPA is presented.
In section~4 the inclusion of collisions is discussed. 
The electron-ion collisions are considered in lowest order
perturbation theory. In this approximation the absorption coefficient
for inverse bremsstrahlung is related to the imaginary part of the
force-force correlation function. 
Finally Section~5 concludes with a discussion of the results and gives
a short outlook.


\setcounter{equation}{0}
\section{Canonical Description in the Hyperplane Formalism}
In this section we demonstrate the derivation of a quantum Hamiltonian
starting from the classical QED Lagrangian. The main issues like gauge
fixing, or canonical quantization on the hyperplane are discussed to
some extend. 

\subsection{The Lagrangian Formulation of the Plasma}
We consider a charge neutral system consisting of two species of
fermions, like for instance electrons (mass $m$, charge $Z_e e$,
$Z_e = -1$) and ions (mass $m_i$, charge $Z_i e$). In particular, we
consider protons ($Z_i = 1$), but the generalization to arbitrary
charged particles is straight forward. 
The standard covariant formulation of a QED plasma is given in terms
of the Lagrangian ${\mathcal L}$
\begin{equation}
\label{QED:LagrPr}
{\mathcal L}'(x)
= {\mathcal L}^{}_{D}(x) + {\mathcal L}^{}_{EM}(x) 
+ {\mathcal L}^{}_{\rm  int}(x) 
\end{equation}
where  ${\mathcal L}^{}_{D}(x)$ is the Dirac part, describing the
fermionic components of the plasma (e.g. electrons and protons),
${\mathcal L}^{}_{EM}(x)$ is the electro-magnetic part and
${\mathcal L}^{}_{\rm int}(x)$ describes the interaction in the
plasma. In standard notation~\cite{Zuber,Gross} we can write the different
terms as
\begin{eqnarray}
\label{DirLagr}
& &
{\mathcal L}^{}_{D}(x)
= \sum_{c=e,i}^{}
\bar\psi_{}^{c}(x)\left({i\over2} \gamma^{\mu}{\difleftright}^{}_{\mu}
- m \right)\psi_{}^{c}(x) ~,
\\[4pt]
& &
\label{EMLagr}
{\mathcal L}^{}_{EM}(x)
= -\frac{1}{4}F_{\mu\nu }(x) F^{\mu\nu }(x) ~,
\\[4pt]
& &
\label{IntLagr}
{\mathcal L}^{}_{\rm int}(x)
= - \sum_{c=e,i}^{} Z_{c}^{}e
j_{\mu}^{c}(x) A^{\mu}(x) ~,
\end{eqnarray}
where
${\difleftright}^{}_{\mu}={\difright}^{}_{\mu}-{\difleft}^{}_{\mu}$.
In the following the electro-magnetic field tensor is taken
in the form
$F^{}_{\mu\nu}=\partial^{}_{\mu} A^{}_{\nu}-\partial^{}_{\nu} A^{}_{\mu}$.
The current density four-vector will be expressed as
$j_{\mu}^{c}=\bar\psi_{}^{c}\gamma_{\mu}^{}\psi_{}^{c}$ with $e<0$.

Additionally to the terms given in Eq.~(\ref{QED:Lagr}) we consider
the influence of an external field $A^{\mu}_{\rm ext}(x)$ on the
system. This external field is not a dynamical field variable, but is
some given function of space and time. This implies that there is no
back reaction mechanism of the system onto $A^{\mu}_{\rm ext}(x)$.
The external field couples to the fermion current and therefore the
complete Lagrangian takes the form
\begin{eqnarray}
\label{QED:Lagr}
& &
{\mathcal L}(x) 
= {\mathcal L}^{\prime}_{}(x) + {\mathcal L}^{}_{\rm  ext}(x),
\\[3pt]
\label{ExtLagr}
& &
{\mathcal L}^{}_{\rm ext}(x)
=  - \sum_{c=e,i}^{} Z_{c}^{}e j_{\mu}^{c}(x) A^{\mu}_{\rm ext}(x).
\end{eqnarray}

\subsection{Gauge fixing}
The Lagrangian given in Eq.~(\ref{QED:Lagr}) is covariant and gauge
invariant. However, as well known, the gauge invariance leads to
non-physical degrees of freedom, which have to be eliminated from the
description. One possible way to perform this elimination is to apply
an additional gauge condition. We will follow this way and use the
Coulomb gauge $\vek{\nabla}\cdot \vek{A}=0$, which is the most natural
choice for Coulomb systems, since the Coulomb interaction appears
naturally in this description. 

The disadvantage of this method, however, is that the explicit
covariance is lost. In particular, after application of some further
approximations it is hard to control, if the final result will be
expressible in covariant form. For this reason it is most advantageous
to use the hyperplane formalism~\cite{Fleming_Rohrlich}, which singles out time-like and
space-like parts in a covariant way. 

Beside the natural appearance of the instantaneous Coulomb interaction
due to the breaking of the explicit covariance by the Coulomb gauge
constraint, there is a second advantage for this method. In view of
laser plasma interactions in general we have to deal with a highly
correlated plasma, which is subject to the external laser field. In
order to describe the correlated plasma initially, the time coordinate
has to be singled out, which leads to the breaking of covariance.
As already mentioned above, in the hyperplane formalism the inclusion
of initial correlation as well as the gauge fixing leading to the
Coulomb interaction can be performed in a manifest covariant manner.
The main issues of this formalism will be discussed in the next section.

\subsection{Introduction of space-like hyperplanes}
A space-like hyperplane $\sigma \equiv\sigma_{n,\tau}^{}$ in Minkowski
space can be characterized by a unit time-like normal
vector $n^{\mu}$ and a scalar parameter $\tau$ which may be interpreted as an
``\,invariant time''. The equation of the hyperplane
$\sigma^{}_{n,\tau}$ reads
 \begin{equation}
 \label{EqPlane}
 x\cdot n=\tau ~, \qquad
n^2=n^{\mu} n^{}_{\mu}=1 ~.
\end{equation}
In the special Lorentz frame where $n^{\mu}=(1,0,0,0)$ and
consequently Eq.~(\ref{EqPlane}) reads $x^0=\tau$ the parameter
$\tau$ coincides with the time variable $t=x^0$.
We will refer to this special frame as the ``instant frame'', since
only here observables are measured at the same instant of time $t$.
Expressing the field variables as functionals of the hyperplane
$\psi[\sigma_{n,\tau}^{}]$ or as functions of $n$ and $\tau$,
The Lagrangian can be expressed on the hyperplane
$\sigma_{n,\tau}^{}$.
The gauge condition is expressed on the plane according to 
\begin{equation}
\label{GaugeCond}
\nabla^{}_{\mu} A^{\mu}_{\trans}=0 ~,
\end{equation}
where the following decomposition of four vectors is used
\begin{equation}
\label{DecompVec}
V^{\mu}=n^{\mu} V^{}_{\longi} + V^{\mu}_{\trans},
\qquad V^{}_{\longi}=n^{}_{\nu}
V^{\nu}, \quad V^{\mu}_{\trans}=\Delta^{\mu}_{\  \nu} V^{\nu} ~,
\end{equation}
\begin{equation}
\label{DecompDer}
\partial^{}_{\mu}
= n^{}_{\mu}\frac{\partial}{\partial\tau}
+ \nabla^{}_{\mu},
\qquad
\nabla^{}_{\mu} = \Delta^{\ \nu}_{\mu} \partial^{}_{\nu}=
\Delta^{\ \nu}_{\mu}\,\frac{\partial}{\partial x^{\nu}_{\trans}}\, ~,
\end{equation}
with the transverse projector $\Delta^{\mu}_{\ \nu}$
\begin{equation}
\label{TransProj}
\Delta^{\mu}_{\ \nu} = \delta^{\mu}_{\ \nu} -n^{\mu} n^{}_{\nu} ~.
\end{equation}
Applying the decomposition~(\ref{DecompVec}) -- (\ref{DecompDer}) to
the Lagrangian~(\ref{QED:Lagr}), we obtain
\begin{eqnarray}
\label{LagrDecomp}
& & {\mathcal L}= - {1\over4} F^{}_{\trans \mu\nu}
F^{\mu\nu}_{\trans} - {1\over2}
\left(
\nabla^{\mu} A^{}_{\longi} - \dot{A}^{\mu}_{\trans}\right) \left(
\nabla^{}_{\mu} A^{}_{\longi} - \dot{A}^{}_{\trans\mu}\right)
\nonumber\\[4pt]
& &
\hspace*{90pt}
{}- \sum_{c} Z_{c}^{} e \Big( 
j^{c}_{\longi} A^{}_{\longi} - j^{c}_{\trans\mu} \Big)
A^{\mu}_{\trans} +{\mathcal L}^{}_{D} + {\mathcal L}^{}_{\rm ext} ~,
\end{eqnarray}
where we have introduced the notation
\begin{equation}
\label{TransEMTens}
F^{\mu\nu}_{\trans}= \nabla^{\mu} A^{\nu}_{\trans} - \nabla^{\nu}
A^{\mu}_{\trans} ~.
\end{equation}
Due to the gauge condition~(\ref{GaugeCond}) we find a constraint
equation, similar to the Poisson equation
\begin{equation}
\label{GenPoisson} \nabla^{}_{\mu} \nabla^{\mu}
A^{}_{\longi} = \sum_{c} Z_{c}^{} e j^{c}_{\longi} ~.
\end{equation}
The solution of Eq.~(\ref{GenPoisson}) is
\begin{equation}
\label{AParall:cl}
{A}^{}_{\longi}(\tau,x^{}_{\trans})= \sum_{c} Z_{c}^{} e
\int_{\sigma^{}_{n}} d\sigma'\,
G(x^{}_{\trans}-x^{\prime}_{\trans})\,{j}^{c}_{\longi}
(\tau,x^{\prime}_{\trans}) ~,
\end{equation}
where the Green function $G(x^{}_{\trans})$ satisfies the equation
\begin{equation}
\label{GrFunc:Eq}
\nabla^{}_{\mu}\nabla^{\mu} G(x^{}_{\trans})=\delta^{3}(x^{}_{\trans})
\end{equation}
and the three-dimensional delta function on a hyperplane
$\sigma^{}_{n}$ defined as
\begin{equation}
\label{DeltaFunc}
\delta^{3}(x^{}_{\trans})=
\int \frac{d^4 p}{(2\pi)^3}\, {\rm e}^{-ip\cdot x}\,
\delta(p\cdot n) ~.
\end{equation}
The solution of Eq.~(\ref{GrFunc:Eq}) for $G(x^{}_{\trans})$ is given by
\begin{equation}
\label{GrFunc}
G(x^{}_{\trans})=
- \int \frac{d^{4}p}{(2\pi)^{3}}\, {\rm e}^{-ip\cdot x}\,
\delta(p\cdot n)\,\frac{1}{p^{2}_{\trans}} ~.
\end{equation}
The variable $A^{}_{\longi}$ can now be eliminated from the Lagrangian
density~(\ref{LagrDecomp}) using Eq.~(\ref{AParall:cl}) imposing
appropriate boundary conditions. 
Then a straightforward  algebra leads to
\begin{eqnarray}
\label{LagrDecomp1}
{\mathcal L} 
= &-& {1\over4} F^{}_{\trans \mu\nu} F^{\mu\nu}_{\trans}
-{1\over2} \dot{A}^{}_{\trans\mu} \dot{A}^{\mu}_{\trans}
- \sum_{c} Z_{c}^{} e j^{c}_{\trans\mu} A^{\mu}_{\trans}
\nonumber\\[4pt]
{}&+&
{\mathcal L}^{}_{D} + {\mathcal L}^{}_{\rm ext}
-{e_{}^{2}\over2}\sum_{c,c'} Z_{c}^{} Z_{c'}^{} 
\int\limits_{\sigma^{}_{n}} d\sigma'\,
j^{c}_{\longi}(\tau,x^{}_{\trans}) 
G(x^{}_{\trans}-x^{\prime}_{\trans})
j^{c'}_{\longi}(\tau,x^{\prime}_{\trans}) ~.
\end{eqnarray}

\subsection{The commutation and anti-commutation relations}
In the canonical quantization scheme, which will be applied here, the 
commutation and anti-commutation relations of the canonical field
operators have to be derived. We follow the method, originated by
Dirac~\cite{Dirac50,Weinberg96}. The momentum $\Pi^{}_{\trans\mu}$
canonical to the electro-magnetic field variable $A_{\trans\mu}^{}$ is
defined as
\begin{equation}
\label{Momenta}
\Pi^{}_{\trans\mu} = \frac{\partial{\mathcal L}}
{\partial \dot{A}^{\mu}_{\trans}} = -\dot{A}^{}_{\trans\mu} ~,
\end{equation}
with the $\tau$-derivative $\dot{A}^{}_{\trans\mu}$.
Similarly, for the fermionic field variables we define the canonical
momenta $\bar\pi_{}^{}$ and $\pi_{}^{}$ according to
\begin{equation}
\label{Fermi:CanMom}
\bar\pi_{}^{}\equiv
\frac{\partial{\mathcal L}_{D}^{}}{\partial \dot{\psi}}
= \frac{i}{2} \bar\psi\gamma_{\longi}^{},
\qquad
\pi_{}^{}\equiv
\frac{\partial{\mathcal L}_{D}^{}}
{\partial \,\dot{\!\bar\psi}} =
- \frac{i}{2} \gamma_{\longi}^{} \psi ~,
\end{equation}
with the decomposition of Dirac's $\gamma$-matrices
\begin{equation}
\hspace*{-25pt}
\label{Gammas:n}
\gamma^{\mu}= n^{\mu} \gamma^{}_{\longi}(n) +
\gamma^{\mu}_{\trans}(n),
\quad
\gamma^{}_{\longi}(n)=n^{}_{\nu} \gamma^{\nu},
\quad
\gamma^{\mu}_{\trans}(n)=\left(\delta^{\mu}_{\ \nu}
- n^{\mu}  n^{}_{\nu} \right)\gamma^{\nu}.
\end{equation}
The dynamical fields and its canonical momenta will now be interpreted
as operators, satisfying commutator and anti-commutator
relations. These relations can be obtained by calculating the Dirac
brackets, which account for constraints, like the gauge-fixing
constraint Eq.~(\ref{GaugeCond}). In Appendix~A we shortly review the
calculation, leading to the following non-vanishing relations
\begin{eqnarray}
\label{Comm:Can}
& &
\left[\hat{A}^{\mu}_{\trans}(\tau, x^{}_{\trans}),
\hat{\Pi}^{\nu}_{\trans}(\tau, x^{\prime}_{\trans})\right]=
ic^{\mu\nu}(x^{}_{\trans} - x^{\prime}_{\trans}) ~,
\\[5pt]
\label{Comm:CanZero}
& &
\left[\hat{A}^{\mu}_{\trans}(\tau,x^{}_{\trans}),
\hat{A}^{\nu}_{\trans}(\tau,x^{\prime}_{\trans})
\right]=\left[\hat{\Pi}^{\mu}_{\trans}(\tau,x^{}_{\trans}),
\hat{\Pi}^{\nu}_{\trans}(\tau,x^{\prime}_{\trans})
\right]=0 ~,
\end{eqnarray}
where
\begin{equation}
\label{DiracDelta}
c^{\mu\nu}(x^{}_{\trans} -x^{\prime}_{\trans}) =
\int \frac{d^{4}p}{(2\pi)^{3}}\, {\rm e}^{-ip\cdot(x-x')}\,
\delta(p\cdot n)
\left[
\Delta^{\mu\nu}-
\frac{p^{\mu}_{\trans} p^{\nu}_{\trans}}{p^{2}_{\trans}}
\right] ~.
\end{equation}
For the Dirac field operators the anti-commutation relations on the
hyperplane take the form
\begin{eqnarray}
\label{Anticomm1}
& &
\bigg\{
\hat{\psi}_{ac}^{} (\tau ,x_{\trans}^{}),
\,\hat{\!\bar\psi}_{\!a'c'}^{} (\tau ,x_{\trans}^{\prime})
\bigg\}
=\left[\gamma^{}_{\longi}(n)\right]^{}_{aa'}
\delta_{c,c'}^{}
\delta_{}^{3}(x_{\trans}^{} - x_{\trans}^{\prime}),
\\[4pt]
\label{Anticomm2}
& &
\bigg\{
\hat{\psi}_{ac}^{} (\tau ,x_{\trans}^{}),
\,\hat{\psi}^{}_{a'c'} (\tau ,x_{\trans}^{\prime})
\bigg\}
= \bigg\{
\hat{\!\bar\psi}_{\!ac}^{} (\tau ,x_{\trans}^{}),
\,\hat{\!\bar\psi}_{\!a'c'} (\tau ,x_{\trans}^{\prime})
\bigg\}
=0,
\end{eqnarray}
where $a,\,a'$ are the spinor indices and $c,\,c'$ denote the
different species.
In the special Lorentz frame where
$x^{\mu}=(t,\vek{r})$ and $n^{\mu}=(1,0,0,0)$,
we have $\gamma^{}_{\longi}=\gamma^{0}$ and
$\delta_{}^{3}(x_{\trans}^{} - x_{\trans}^{\prime})=\delta(\vek{r}-\vek{r}')$,
so that
Eq.~(\ref{Anticomm1}) reduces to the well-known anticommutation
relation for the quantized Dirac field.

\subsection{The Hamiltonian}
The quantum Hamiltonian can be constructed by a Legendre
transformation, defined on the hyperplane as
\begin{equation}
\label{LegendTrans}
H(n) =
\int_{\sigma_{n,\tau}^{}}^{}d\sigma \, \bigg\{
\Pi^{}_{\trans\mu}\dot{A}_{\trans}^{\mu}
+ \bar\pi \dot{\psi}
+ \,\dot{\!\bar\psi} \pi
- \mathcal{L}^{} \bigg\} ~,
\end{equation}
where ${\mathcal L}$ is given by Eq.~(\ref{LagrDecomp1}).
The Hamiltonian can be written in the form
\begin{equation}
\label{TotalHam:Q} \hat H^{\tau}(n)
= \hat H^{}_{D}(n)+\hat H^{}_{EM}(n) + \hat H^{}_{\rm int}(n) 
+ \hat H^{\tau}_{\rm ext}(n) ~,
\end{equation}
where $\hat H^{}_{D}(n)$
and $\hat H^{}_{EM}(n)$ are the Hamiltonians for free fermions and  the
polarization EM field respectively, $\hat H^{}_{\rm int}(n)$ is the
interaction term, and $\hat H^{\tau}_{\rm ext}(n)$
describes the external EM
field effects. In the Schr\"odinger picture the explicit
expressions for these terms are
\begin{eqnarray}
\label{DirHam:Q}
& & \hat{H}^{}_{D}(n)= 
\sum_{c}\int_{\sigma^{}_{n}} d\sigma\,
\,\hat{\!\bar \psi}_{c} \left( -\frac{i}{2} \gamma^{\mu}_{\trans}(n)
\nablaleftright^{}_{\mu} +
m_{c}^{} \right)\hat{\psi}_{c} ~,
\\[4pt]
\label{EMHam:Q}
& &
\hat{H}^{}_{EM}(n)=
\int_{\sigma^{}_{n}}  d\sigma\,
\left(
\frac{1}{4} \hat{F}_{\trans \mu\nu}^{} \hat{F}_{\trans}^{\mu\nu}
- \frac{1}{2} \hat{\Pi}_{\trans \mu}^{}\hat{\Pi}_{\trans}^{\mu}
\right) ~,
\\[4pt]
\label{IntHam:Q}
& &
\hat{H}^{}_{\rm int}(n)
= \frac{e_{}^{2}}{2} \sum_{cc'}Z_{c}^{}Z_{c'}^{}
\int_{\sigma^{}_{n}} d \sigma \int_{\sigma^{}_{n}}
d \sigma_{}^{\prime} \,
\,\hat{\!j}_{\!\longi}^{c}(x_{\trans}^{})
G(x_{\trans}^{} - x_{\trans}^{\prime} )
\,\hat{\!j}_{\!\longi}^{c'}(x_{\trans}^{\prime}) 
\nonumber \\[4pt]
& & \qquad \qquad
+ \sum_{c}
\int_{\sigma^{}_{n}}  d\sigma\,
Z_{c}^{}e\,\,\hat{\!j}^{c}_{\!\trans \mu} \hat{A}_{\trans}^{\mu}~,
\\[4pt]
\label{ExtHam:Q}
& &
\hat{H}^{\tau}_{\rm ext}(n)
= \sum_{c} \int_{\sigma^{}_{n}} d\sigma\,
Z_{c}^{}e\,\,\hat{\!j}^{c}_{\!\mu}(x^{}_{\trans}) {A}^{\mu}_{\rm ext}(\tau,
x^{}_{\trans}) ~.
\end{eqnarray}
The field strength tensor $\hat{F}_{\!\trans \mu\nu}^{}$ and the
transverse field operators $\hat{A}_{\!\trans}^{\mu}$ and 
$\hat{\Pi}_{\!\trans}^{\mu}$ in Eq.~(\ref{EMHam:Q}) are defined
according to their classical relations (\ref{DecompVec}) and 
(\ref{TransEMTens}).
The longitudinal part $\hat{A}^{}_{\longi}$ has been eliminated
in the interaction Hamiltonian~(\ref{IntHam:Q}) by the operator
version of Eq.~(\ref{GenPoisson}).
In Eqs.~(\ref{DirHam:Q})\,--\,(\ref{ExtHam:Q})
normal ordering in operators
is implied. The self-energy contribution to
the last term in Eq.~(\ref{IntHam:Q}) is
omitted, so that the product
$\Norder\,\hat{\!j}^{c}_{\!\longi}(x^{}_{\trans})\Norder\,\Norder
\,\hat{\!j}^{c}_{\!\longi}(x'_{\trans})\Norder$
is understood. 
The generalization of the Hamiltonian to a many-component case is obvious.

It should be mentioned, that the same expressions for the Hamiltonian
Eq.(\ref{DirHam:Q}) -- (\ref{ExtHam:Q})
can be obtained from the symmetrized energy-momentum tensor, the so
called Belinfante tensor $T_{\mu\nu}^{}$~\cite{Belinfante39} via
\begin{equation}
\label{HamOnPlane:cl}
H(n)=P^{}_{\mu}n^{\mu}\equiv
\int_{\sigma^{}_{n,\tau}} d\sigma\,
n^{\mu}T^{}_{\mu\nu}n^{\nu} ~.
\end{equation}

\subsection{The relativistic von Neumann equation}
As already noted, the state of the system
$|\Psi[\sigma^{}_{n,\tau}]\rangle$ is taken as a functional of
$\sigma^{}_{n,\tau}$. 
A specific frame of reference can now be related to a
family of space-like hyperplanes with a fixed normal vector $n$.
The relation between different frames of reference is given by a
homogeneous Lorentz transformation $\Lambda$ 
\begin{equation}
\label{LorTrf}
\sigma_{n,\tau}^{}\to
\sigma_{n',\tau}^{}=\Lambda\sigma_{n,\tau}^{}: \quad x\to x'=\Lambda x ~.
\end{equation}
Eq.~(\ref{LorTrf}) means that 
$x_{\mu}^{}n_{}^{\mu} = x_{\mu}^{\prime}n_{}^{\prime\mu} = \tau$, i.e.
$x$ is located at $\sigma$ and $x'$ at $\sigma'$.

With a unitary representation of the homogeneous Lorentz group
$U(\Lambda)$ state vectors on different planes are related by~\cite{Schweber61}
\begin{equation}
\label{TrfStVec}
U(\Lambda)\left|\Psi[\Lambda\sigma]\right\rangle=
\left|\Psi[\sigma]\right\rangle ~.
\end{equation}
Having $n_{\mu}^{}$ fixed, the evolution of the state in this frame of
reference is governed by a representation $U(a)$ of time-like translations
$a_{\mu}^{}$ in the direction of the normal $n_{\mu}^{}$. The
generator of this transformation is $\hat P^{\mu}$, the energy-momentum
vector. $U(a)$ is given by
\begin{equation}
\label{Transl}
U(a)=\exp\left\{i\hat P^{}_{\mu} a^{\mu}\right\} ~.
\end{equation}
Writing the states as functions of $n$ and $\tau$, we find for an
infinitesimal time-like translation $a^{\mu}=n^{\mu}\,\delta\tau$ 
\begin{equation}
\label{InfTrans}
\left|\Psi(n,\tau+\delta\tau)\right\rangle 
+ i\delta\tau \left(\hat P^{}_{\mu} n^{\mu}\right)
\left|\Psi(n,\tau)\right\rangle
= \left|\Psi(n,\tau)\right\rangle ~,
\end{equation}
from which we obtain the relativistic Schr\"odinger  equation
\begin{equation}
\label{SchrEq}
i\frac{\partial}{\partial\tau}
\left|\Psi(n,\tau)\right\rangle = \hat H(n) \left|\Psi(n,\tau)\right\rangle
\end{equation}
with the Hamiltonian on the hyperplane given by
\begin{equation}
\label{HamOnPlane}
\hat H(n)=\hat P^{}_{\mu} n^{\mu} ~.
\end{equation}
In the presence of a
prescribed external field, the energy-momentum vector and,
consequently, the
Hamiltonian $\hat H^{\tau}(n)$ can depend explicitly on $\tau$. Combining
Eq.~(\ref{SchrEq}) with the adjoint equation for the bra-vector, one finds
that the statistical operator $\varrho(n,\tau)$ for a mixed quantum
ensemble obeys the equation
\begin{equation}
\label{VonNEq}
\frac{\partial\varrho(n,\tau)}{\partial\tau}
- i\left[\varrho(n,\tau),\hat H^{\tau}(n)\right]=0 ~,
\end{equation}
which is analogous to
the non-relativistic von Neumann equation.

In order to solve Eq.~(\ref{VonNEq}), some boundary condition have to
be imposed on the statistical operator. The standard boundary condition in
kinetic theory is Bogoliubov's boundary condition of weakening of initial
correlations which implies the uncoupling of all correlation functions to
one-particle density matrices in the distant past, i.e., for
$\tau\to -\infty$.
In the scheme
developed by Zubarev (see, e.g.,~\cite{ZubMorRoep1}),
such boundary conditions can be included by using
instead of Eq.~(\ref{VonNEq})
the equation with an infinitesimally small source term
 \begin{equation}
 \label{ZubEq}
\frac{\partial\varrho(n,\tau)}{\partial\tau} -
i\left[\varrho(n,\tau),\hat{H}^{\tau}(n)\right]=
-\eta\left\{
\varrho(n,\tau)-\varrho^{}_{\rm rel}(n,\tau)
\right\},
 \end{equation}
where $\eta\to +0$ after the calculation of averages.
Here $\varrho^{}_{\rm rel}(n,\tau)$ is the so-called
\emph{relevant statistical operator\/} which describes a Gibbs state
for some given non-equilibrium state variables.
In QED kinetics these variables are
the Wigner function and the photon density
matrix.
In general we will call these non-equilibrium state variables the
relevant operators $\hat{B}_{\ell}^{\mu}$. In the next section we will
search for solutions of Eq.~(\ref{ZubEq}) up to terms linear in
$\hat{B}_{\ell}^{\mu}$. 

\setcounter{equation}{0}
\section{Linear Response Theory}
Using the results of the last section, it is possible to construct a
kinetic theory on the hyperplanes, which was extensively discussed
in~\cite{hoell1}.

Here we show how the non-equilibrium statistical operator for the case
of small deviations from the equilibrium distribution can be
constructed. This can be done by solving Eq.~(\ref{ZubEq}) with a
appropriate choice for the relevant distribution.
This treatment, known as generalized linear response, was successfully
applied in different studies~\cite{Gerd}.

Further for simplicity we consider the plasma in the adiabatic
approximation, where the dynamics of the positively charged component
of the plasma is frozen.

\subsection{Fluctuations near equilibrium}
For a given relevant distribution $\varrho_{rel}^{}$, a formal
solution of the Zubarev equation~(\ref{ZubEq}) is given by 
\begin{equation}
\label{StOp1}
\varrho(n,\tau)=
\eta \int\limits^{\tau}_{-\infty} d\tau'\,
{\rm e}^{-\varepsilon(\tau -\tau')}\,
U(\tau,\tau')\,{\varrho}^{}_{\rm rel}(n,\tau')\,
U^{\dagger}(\tau,\tau') ~,
\end{equation}
where the evolution operator can be written as the ordered exponent
\begin{equation}
\label{EvOp}
U(\tau,\tau') =
T^{}_{\tau}\,\exp\left\{-i \int\limits^{\tau}_{\tau'}
\hat H^{\bar\tau}(n)\,d\bar{\tau}  \right\} ~.
\end{equation}
After partial integration, the expression~(\ref{StOp1}) becomes
\begin{equation}
\label{StOp2}
\varrho(n,\tau)=
\varrho^{}_{\rm rel}(n,\tau) +
\Delta{\varrho}(n,\tau) ~,
\end{equation}
\begin{eqnarray}
\label{DeltaRho}
& &
\hspace*{-20pt}
\Delta{\varrho}(n,\tau)=
-\int\limits^{\tau}_{-\infty} d\tau'\,
{\rm e}^{-\eta(\tau -\tau')}
\nonumber\\[8pt]
& &
\hspace*{40pt}
{\times}\,
U(\tau,\tau')
\left\{
\frac{\partial {\varrho}^{}_{\rm rel}(n,\tau')}{\partial\tau'}
-i\left[{\varrho}^{}_{\rm rel}(n,\tau'),
\hat{H}^{\tau'}(n)\right]
\right\}
U^{\dagger}(\tau,\tau') ~.
\end{eqnarray}

The Hamiltonian $\hat{H}$, as defined in the Eqs.~(\ref{TotalHam:Q}) --
(\ref{ExtHam:Q}), is rewritten in the adiabatic approximation 
in decomposed form according to 
\begin{eqnarray}
\label{TotHam:Ad}
& &
\hat{H}(n) = \hat{H}_{s}^{}(n) + \hat{H}_{ext}^{}(n) ~,
\\[4pt]
\label{SysHam:Ad}
& &
\hat{H}_{s}^{}(n) = \hat{H}_{D}^{}(n)\Big|_{c=e}^{} + \hat{H}_{EM}^{}(n) +
\hat{H}_{rad}^{}(n)\Big|_{c=e}^{} +  \hat{H}_{int}^{}(n) ~.
\end{eqnarray}
In (\ref{TotHam:Ad}) -- (\ref{SysHam:Ad}) the system part of the
Hamiltonian $\hat{H}_{s}^{}$ is decomposed into the kinetic parts
$\hat{H}_{D}^{}$ and $H_{EM}^{}$ and some part describing the
collisions $\hat{H}_{int}^{}$ and $\hat{H}_{rad}^{}$, which will be
treated within perturbation theory. 
The different parts of the Hamiltonian are taken from
Eqs.~(\ref{TotalHam:Q}) -- (\ref{ExtHam:Q})
\begin{eqnarray}
\label{DirHam:Ad}
& & \hat{H}^{}_{D}(n)
=  \int_{\sigma^{}_{n}} d\sigma\,
\,\hat{\!\bar \psi}_{e} \left( -\frac{i}{2} \gamma^{\mu}_{\trans}(n)
\nablaleftright^{}_{\mu} + m_{e}^{} \right)\hat{\psi}_{e} ~,
\\[4pt]
\label{EMHam:Ad}
& &
\hat{H}^{}_{EM}(n)=
\int_{\sigma^{}_{n}}  d\sigma\,
\left(
\frac{1}{4} \hat{F}_{\trans \mu\nu}^{} \hat{F}_{\trans}^{\mu\nu}
- \frac{1}{2} \hat{\Pi}_{\trans \mu}^{}\hat{\Pi}_{\trans}^{\mu}
\right) ~,
\\[4pt]
\label{RadHam:Ad}
& & 
\hat{H}^{}_{rad}(n)
= - e \int_{\sigma^{}_{n}}  d\sigma\,
\,\,\hat{\!j}^{e}_{\!\trans \mu}(x_{\trans}^{}) 
\hat{A}_{\trans}^{\mu}(x_{\trans}^{}) ~,
\\[4pt]
\label{InteractionHam:Ad}
& & 
\hat{H}^{}_{int}(n)
= - e
\int_{\sigma^{}_{n}} d \sigma \,
\left(
\,\hat{\!j}_{\!\longi}^{e}(x_{\trans}^{})
A_{\longi\,}^{ion}(x_{\trans}^{})
+ \,\hat{\!j}_{\!\trans}^{e\mu}(x_{\trans}^{})
A_{\trans\,\mu}^{ion}(x_{\trans}^{})
\right) ~,
\\[4pt]
\label{ExtHam:Ad}
& &
\hat{H}^{\tau}_{\rm ext}(n)
= - e \int_{\sigma^{}_{n}} d\sigma\,
\,\,\hat{\!j}^{e}_{\!\mu}(x^{}_{\trans}) {A}^{\mu}_{\rm ext}(\tau,
x^{}_{\trans}) ~.
\end{eqnarray}
In the following we will drop the index ``$e$'' and the
electron spinors are denoted by $\,\hat{\!\bar\psi}_{}$ and
$\hat{\psi}_{}$.

The radiation term, Eq.~(\ref{RadHam:Ad}), is important for the
description of photon emission from the plasma or for photon
scattering in the plasma.
As an application we will focus to the absorption of a classical
electro-magnetic wave by a relativistic plasma in the next section,
where the radiation term can be neglected.

In what follows, the non-equilibrium state of the system must be
specified. We use Zubarev's method~\cite{ZubMorRoep1} of a
non-equilibrium statistical ensemble in linear response. 
The so called relevant statistical operator ${\varrho}^{}_{\rm rel}(n,\tau)$
describes a generalized Gibbs distribution, which
characterizes the initial non-equilibrium state of our system. In the
linear response regime we will only consider small fluctuations from
the equilibrium. 
The relevant statistical operator can be written in the form 
\begin{eqnarray}
\label{Rho:Relevant} 
\varrho_{{\rm rel}}^{}(n,\tau) 
&=& 
Z_{{\rm rel}}^{-1}(\beta_{}^{} , \nu , \phi_{\mu}^{\ell};\tau) 
\exp \Bigg\{ - \beta_{}^{} \bigg[ n_{}^{\mu} \hat{P}_{\mu}^{} 
- \nu_{}^{} \hat{Q}_{}^{}
\nonumber \\
& &
- \int_{\sigma_{n}^{}} d\sigma ~ \sum_{\ell} \phi_{\mu}^{\ell}(x_{\trans}^{};\tau) 
\hat{B}_{\ell}^{\mu}(x_{\trans}^{}) \bigg] \Bigg\} ~,
\end{eqnarray}
where the relevant observables
$\hat{B}_{\ell}^{\mu}(x_{\trans}^{})$ define the non-equilibrium state
and will be treated as small quantities.
It should be mentioned, that the $\tau$-de\-pen\-dence is carried
completely by the set of Lagrange multipliers 
$\phi_{\mu}^{\ell}(x_{\trans}^{};\tau)$.
The first two terms in the exponential of Eq.~(\ref{Rho:Relevant})
describe the generalization of the Gibbsian distribution
$\varrho_{0}^{}(n)$ of the grand canonical ensemble
\begin{eqnarray}
\label{Rho:Eq}
\varrho_{0}^{}(n) 
= Z_{0}^{-1}(\beta_{}^{} , \nu ; n) 
\exp \bigg\{ - \beta_{}^{} \Big[ n_{}^{\mu} \hat{P}_{\mu}^{} 
- \nu \hat{Q} \Big] \bigg\} ~,
\end{eqnarray}
where
\begin{eqnarray}
\label{H0}
n_{}^{\mu} \hat{P}_{\mu}^{} = \hat H_{s}^{}
~,\qquad
\hat{Q} = n_{\mu}^{} \,\hat{\!j}_{}^{\mu} ~.
\end{eqnarray}
The equilibrium statistical operator does only depend on the additive
integrals of motion $\hat{P}_{\mu}^{}$, $\hat{Q}$.

In the following we demonstrate how correlation functions can
be derived in linear response within the hyperplane formalism.
This means, that all expressions will be approximated to first order
in quantities, describing the deviation from the equilibrium. 

Expressing the relevant part of the statistical operator
Eq.~(\ref{Rho:Relevant}) in linear response, as demonstrated in
Appendix~B, we find in Fourier representation
\begin{eqnarray}
\label{Rho:RelLin1F}
\varrho_{{\rm rel}}^{}(n,\tau) 
&=& \varrho_{0}^{}(n) + \beta\, {\rm e}_{}^{-i k_{\longi}^{}\cdot \tau} 
\int_{0}^{1} dz \,
\sum_{\ell} \phi_{\mu}^{\ell}(k) \,
\hat{B}_{\ell}^{\dagger\mu}(k_{\trans}^{},iz\beta) 
\varrho_{0}^{}(n) ~.
\end{eqnarray}
A similar calculation leads to the irrelevant part of the statistical
operator in linear response as (see Appendix~B for details)
\begin{eqnarray}
\label{Rho:IrrLin2F}
\Delta \varrho_{}^{}(n,\tau)
&=& \beta\, {\rm e}_{}^{-ik_{\longi}^{}\tau} \int_{0}^{\infty} d\tilde{\tau}\, 
{\rm e}_{}^{i(k_{\longi}^{}+i\eta) \tilde{\tau}}
\int_{0}^{1} dz
\nonumber \\[2mm]
& &
\bigg\{ \sum_{\ell} \Big[
\!\!\!\dot{\,\,\,\hat{B}_{\ell}^{\dagger\mu}}(-\tilde{\tau}+iz\beta)
-ik_{\longi}^{} \hat{B}_{\ell}^{\dagger\mu}(-\tilde{\tau}+iz\beta)
\Big] \phi_{\mu}^{\ell}(k)
\nonumber \\[2mm]
& &
+ A_{{\rm ext}}^{\mu}(k) 
\,\dot{\hat{\!j}}_{\!\mu}^{} (-\tilde{\tau} + iz\beta)
\bigg\} \varrho_{0}^{}(n) ~.
\end{eqnarray}
By definition, the mean values of the relevant observables are some
prescribed functions of space and time and are obtained by averaging
with the relevant statistical operator only. 
The self-consistency relations, by which the Lagrange multipliers are
determined, can therefore be expressed as 
\begin{eqnarray}
\label{S:ConsRel}
{\rm Tr} \left\{ \hat{B}_{\ell}^{\mu}(x_{\trans}^{}) 
\Delta \varrho_{}^{} (n,\tau ) \right\} = 0 ~.
\end{eqnarray}
Multiplying Eq.~(\ref{Rho:IrrLin2F}) by $\hat{B}_{\ell'}^{\nu}$, taking 
the trace and rearranging the indices we find the response equation
\begin{equation}
\label{S:ConsRel1}
-\Big\langle \hat{B}_{\mu}^{\ell} \,;\,
\dot{\hat{\!j}}_{\!\nu}^{} \Big\rangle_{\!k_{\longi}^{}+i\eta}^{}
A_{{\rm ext}}^{\nu}
= \sum_{\ell'} \Big\langle \hat{B}_{\mu}^{\ell} \,;
\big(\!\!\dot{\,\,\hat{B}_{\nu}^{\ell'}}
+ i k_{\longi}^{} \hat{B}_{\nu}^{\ell'} \big)
\Big\rangle_{\!k_{\longi}^{}+i\eta}^{} \phi_{\ell'}^{\nu} ~.
\end{equation}
In Eq~(\ref{S:ConsRel1}) the correlation functions
$(\hat{A}\, ; \hat{B})$ and its Laplace transforms 
$\langle \hat{A}\, ; \hat{B} \rangle_{\eta}^{}$ are defined 
according to
\begin{eqnarray}
\label{Def:TCorrFns1}
\Big(\hat{A}\, ; \hat{B}\Big) 
&=& \int_{0}^{1} dz\, {\rm Tr} \left\{ \hat{A}(-i\beta z)
  \hat{B}_{}^{\dagger} \varrho_{0}^{}\right\}
= \int_{0}^{1} dz\, {\rm Tr} \left\{ \hat{A}
  \hat{B}_{}^{\dagger}(i\beta z) \varrho_{0}^{}\right\} ~,
\\[4pt]
\label{Def:LaplTCorrFns1}
\Big\langle \hat{A}\, ; \hat{B} \Big\rangle_{\!\eta}^{} 
&=& \int_{0}^{\infty} d\tilde{\tau}\; {\rm e}_{}^{i\eta\tilde{\tau}}
\Big(\hat{A}(\tilde{\tau})\, ; \hat{B}\Big)
= \int_{0}^{\infty} d\tilde{\tau}\; {\rm e}_{}^{i\eta\tilde{\tau}}
\left(\hat{A}\, ; \hat{B}(-\tilde{\tau}) \right) ~.
\end{eqnarray}
Making use of Eqs.~(\ref{Rho:RelLin1F}) and (\ref{Rho:IrrLin2F}) local
observables can be expressed in linear response, in particular the 
induced current 
$j_{{\rm ind}}^{\mu} = \delta\langle \,\hat{\!j}_{}^{\mu} \rangle$,
is written as 
\begin{math}
\left[
\delta\langle \hat{O}_{\mu}^{} \rangle_{}^{\tau}
= \delta\langle \hat{O}_{\mu}^{} \rangle_{}^{k_{\longi}^{}} 
\exp\{-ik_{\longi}\tau\} \right]
\end{math}
\begin{eqnarray}
\label{j:DevEquil1}
\delta\Big\langle \,\hat{\!j}_{\mu}^{} \Big\rangle_{}^{k_{\longi}^{}} 
&=& \beta \sum_{\ell} \bigg\{
\Big( \,\hat{\!j}_{\mu}^{}\, ; \hat{B}_{\nu}^{\ell} \Big)
- \Big\langle \,\hat{\!j}_{\mu}^{}\, ;
\!\!\dot{\,\,\hat{B}_{\nu}^{\ell}} \Big\rangle_{\!k_{\longi}^{}+i\eta}^{}
+ ik_{\longi}^{} \Big\langle \,\hat{\!j}_{\mu}^{}\, ; 
\hat{B}_{\nu}^{\ell} \Big\rangle_{\!k_{\longi}^{}+i\eta}^{}
\bigg\} \phi_{\ell}^{\nu}
\nonumber\\[4pt]
& &
- \beta  \Big\langle \,\hat{\!j}_{\mu}^{} \,;\,
\dot{\hat{\!j}}_{\!\nu}^{}
\Big\rangle_{\!k_{\longi}^{}+i\eta}^{}A_{{\rm ext}}^{\nu}  ~.
\end{eqnarray}
Eq.~(\ref{j:DevEquil1}) describes the response of the system due to
weak perturbations $A_{{\rm ext}}^{\mu}$. The Lagrange multipliers can 
be eliminated from Eq.~(\ref{j:DevEquil1}) by making use the
self-consistency relation~(\ref{S:ConsRel1}).

If the induced current can be represented by a linear
combination of the relevant operators 
\begin{math}
\,\hat{\!j}_{\!\mu}^{} =
\sum_{\ell'}a_{\ell'}^{}\hat{B}_{\mu}^{\ell'},
\end{math} 
Eq.~(\ref{j:DevEquil1}) is simplified as
\begin{eqnarray}
\label{j:DevEqS1}
\delta\Big\langle \,\hat{\!j}_{\mu}^{} \Big\rangle_{}^{k_{\longi}^{}} 
&=& \beta \sum_{\ell\ell'} a_{\ell'}^{}
\Big( \hat{B}_{\mu}^{\ell'}\, ; \hat{B}_{\nu}^{\ell} \Big)
\phi_{\ell}^{\nu}
= \beta \sum_{\ell}
\Big( \,\hat{\!j}_{\mu}^{}\, ; \hat{B}_{\nu}^{\ell} \Big)
\phi_{\ell}^{\nu} ~,
\end{eqnarray}
where the remaining terms in Eq.~(\ref{j:DevEquil1}) are canceled due
to the self-consistency relation~(\ref{S:ConsRel1}).
The Lagrange multipliers can be eliminated in Eq.~(\ref{j:DevEqS1})
by multiplying Eq.~(\ref{S:ConsRel1}) by $a_{\ell}^{}$ and
summing over $\ell$. Further partial integration of correlation functions
\begin{eqnarray}
\label{Rel1:CorrFunc}
iz\Big\langle \hat{A}\, ; \hat{B}\Big\rangle_{z}^{}
+ \Big( \hat{A}\, ; \hat{B} \Big)
= \Big\langle 
\hat{A}\, ; 
\,\dot{\hat{\!B}} 
\Big\rangle_{z}^{}
= - \Big\langle 
\,\,\dot{\hat{\!\!A}}\, ; 
\hat{B} 
\Big\rangle_{z}^{}
\end{eqnarray}
can be applied and finally we obtain
\begin{eqnarray}
\label{S:ConsRel2}
-\sum_{\ell} a_{\ell}^{} \Big\langle \hat{B}_{\mu}^{\ell} \,;\,
\dot{\hat{\!j}}_{\!\nu}^{} \Big\rangle_{\!k_{\longi}^{}+i\eta}^{}
A_{{\rm ext}}^{\nu}
= \sum_{\ell\ell'} a_{\ell}^{}\Big( \hat{B}_{\mu}^{\ell} \,;
\hat{B}_{\nu}^{\ell'}\Big) \phi_{\ell'}^{\nu} ~.
\end{eqnarray}

Now Eq.~(\ref{S:ConsRel2}) can be plugged into Eq.~(\ref{j:DevEqS1})
and we have the induced current expressed in terms of the 
external field within the linear response approximation
\begin{eqnarray}
\label{j:DevEqS2}
\delta\Big\langle \,\hat{\!j}_{\mu}^{} \Big\rangle_{}^{k_{\longi}^{}} 
&=& - \beta \Big\langle \,\hat{\!j}_{\mu}^{}\, 
; \,\dot{\hat{\!j}}_{\nu}^{} \Big\rangle_{\!k_{\longi}^{}+i\eta}^{}
A_{{\rm ext}}^{\nu}
\\[4pt]
&=& 
\label{j:DevEqS3}
-\beta\bigg\{ i k_{\longi}^{}
\Big\langle \,\hat{\!j}_{\mu}^{}\, 
; \,\hat{\!j}_{\nu}^{} \Big\rangle_{\!k_{\longi}^{}+i\eta}^{}
+ \Big( \,\hat{\!j}_{\mu}^{}\, 
; \,\hat{\!j}_{\nu}^{} \Big)
\bigg\} A_{{\rm ext}}^{\nu} ~.
\end{eqnarray}
In the second line again partial integration [Eq.~(\ref{Rel1:CorrFunc})]
was applied.

The susceptibility tensor $\chi_{\mu\nu}^{}$, describing the response
of the system to an external perturbation, is defined as
\begin{eqnarray}
\label{DefSuscep1}
j_{\mu}^{{\rm ind}} (k)
= \chi_{\mu\nu}^{} (k) A_{{\rm ext}}^{\nu}(k)
\end{eqnarray}
and can be read from the equations~(\ref{j:DevEqS2}) and (\ref{j:DevEqS3})
\begin{eqnarray}
\label{Suscept1}
\chi_{\mu\nu}^{} = - \beta \Big\langle \,\hat{\!j}_{\!\mu}^{} \,;\,
\dot{\hat{\!j}}_{\!\nu}^{} \Big\rangle_{\!k_{\longi}^{}+i\eta}^{} 
= - \beta \Big( \,\hat{\!j}_{\!\mu}^{} \,;\,\hat{\!j}_{\!\nu}^{}\Big)
- i \beta k_{\longi}^{} \Big\langle \,\hat{\!j}_{\!\mu}^{} \,;\,
\,\hat{\!j}_{\!\nu}^{} \Big\rangle_{\!k_{\longi}^{}+i\eta}^{}  ~.
\end{eqnarray}
Eq.~(\ref{Suscept1}) gives the susceptibility tensor in terms of the
current-current or the current-force correlation function.  
The major task to proceed is to evaluate these correlation
functions within certain approximations. We will show how perturbation
theory can be applied. The simplest evaluation of the correlation
function is given in the RPA approximation, which will be considered
in the next section. As a next step we demonstrate how collisions can
be included.

Since we are working in the adiabatic approximation, we chose a
reference frame in which the the ions are at rest. This means that it is
most convenient to use the instant frame formulation 
($n_{\mu}^{} = (1,0,0,0)$).

\subsection{The RPA-result for the correlation function}
The susceptibility tensor~(\ref{Suscept1}) can be most easily
calculated in the RPA-approximation. 
In this approximation the force operator $\,\dot{\hat{\!j}}_{\nu}^{}$
in Eq.~(\ref{Suscept1}) is calculated with the Dirac part in the
Hamiltonian only.

It is convenient to perform the calculation using the plane wave
expansion for the field operators $\hat{\psi}$ and 
$\,\hat{\!\bar{\psi}}$ according to
\begin{eqnarray}
\label{PWaveExpans1}
\hat{\psi}(x) &=& \int \frac{d_{}^{3}p}{(2\pi\hbar)_{}^{3/2}} \,
\sqrt{\frac{m}{E_p}} \sum_s \bigg(
\hat{b}_{ps}^{} u(p,s){\rm e}_{}^{-\frac{i}{\hbar}p\cdot x}
+ \hat{d}_{ps}^{\dagger} v(p,s) {\rm e}_{}^{+\frac{i}{\hbar}p\cdot x}
\bigg) ~,
\\[3pt]
\label{PWaveExpans2}
\,\hat{\!\bar{\psi}}(x) &=& \int \frac{d_{}^{3}p'}{(2\pi\hbar)_{}^{3/2}} \,
\sqrt{\frac{m}{E_{p'}}}
\sum_{s'}^{} \bigg(
\hat{d}_{p's'}^{} \bar{v}(p',s'){\rm e}_{}^{-\frac{i}{\hbar}p'\cdot x}
+ \hat{b}_{p's'}^{\dagger} \bar{u}(p',s') 
{\rm e}_{}^{+\frac{i}{\hbar}p'\cdot x} \bigg)  ~.
\nonumber\\
\end{eqnarray}
We use the four-dimensional scalar product and the mass on-shell
condition
\begin{eqnarray}
\label{ScalProd:Pl}
x\cdot p = p_{0}^{}\cdot t 
- \vek{x} \cdot \vek{p} \qquad , \qquad
p_{0}^{} = E_{p}^{} = +\sqrt{\vek{p}_{}^{2} + m_{}^{2}} ~.
\end{eqnarray} 
The operators $\hat{b}$, $\hat{b_{}^{\dagger}}$, $\hat{d}$ and
$\hat{d_{}^{\dagger}}$ are the electron
creation and annihilation operators and the corresponding antiparticle
creation and annihilation operators.
The current operator can be expressed in terms of these operators 
\begin{eqnarray}
\label{TotElCurr}
\,\hat{\!j}_{\!\mu}^{} (\vek{k})
&=& e \int d_{}^{3}x\; \Norder \hat{\bar{\psi}}_{}^{}(x)
(\gamma_{\mu}^{})_{}^{} \hat{\psi}_{}^{}(x)\Norder \,
{\rm e}_{}^{-i\vekind{k}\cdot\vekind{x}}
\nonumber \\[2pt]
&=& e \int \frac{d_{}^{3}p}{(2\pi\hbar)_{}^{3}} \;
\sqrt{\frac{m}{E_p}} \,
\sum_{ss'}^{} 
\Bigg\{
\nonumber\\[3pt]
& & ~~\hspace{-1cm}                                        
\sqrt{\frac{m}{E_{p+\hbar k}}} \,
\Norder \hat{d}_{p+\hbar k,s'}^{} \hat{d}_{p,s}^{\dagger} \Norder \,
\Big( \bar{v}(p+\hbar k,s') \gamma_{\mu}^{} v(p,s) \Big) \,
{\rm e}_{}^{\frac{i}{\hbar}(E_{p}^{}-E_{p+\hbar k}^{})t}
\nonumber\\[3pt]
& & \hspace{-1cm} +                                        
\sqrt{\frac{m}{E_{p-\hbar k}}} \,
\Norder 
\hat{b}_{p-\hbar k,s'}^{\dagger} \hat{b}_{p,s}^{} \Norder \,
\Big( \bar{u}(p-\hbar k,s') \gamma_{\mu}^{} u(p,s) \Big) \,
{\rm e}_{}^{-\frac{i}{\hbar}(E_{p}^{}-E_{p-\hbar k}^{})t}
\nonumber\\[3pt]
& & \hspace{-1cm} +                                        
\sqrt{\frac{m}{E_{p-\hbar k}}} \,
\Norder 
\hat{d}_{-p+\hbar k,s'}^{} \hat{b}_{p,s}^{} \Norder \,
\Big( \bar{v}(-p+\hbar k,s') \gamma_{\mu}^{} u(p,s) \Big) \,
{\rm e}_{}^{-\frac{i}{\hbar}(E_{p}^{}+E_{p-\hbar k}^{})t}
\nonumber\\[3pt]
& & \hspace{-1cm} +                                        
\sqrt{\frac{m}{E_{p+\hbar k}}} \,
\Norder 
\hat{b}_{-p-\hbar k,s'}^{\dagger} \hat{d}_{p,s}^{\dagger} \Norder \,
\Big( \bar{u}(-p-\hbar k,s') \gamma_{\mu}^{} v(p,s) \Big) \,
{\rm e}_{}^{\frac{i}{\hbar}(E_{p}^{}+E_{p+\hbar k}^{})t}
\Bigg\} .
\end{eqnarray}
The result~(\ref{TotElCurr}) was obtained after integration over
$\vek{x}$ and $\vek{p}'$.
 
In a similar way the Dirac Hamiltonian is written in the plane-wave
representation 
\begin{eqnarray}
\label{DirHam3}
\hat{H}_{D}^{} &=& \int d_{}^{3} p \; 
\sum_{s}^{} \, E_{p}^{}\Big(
\hat{d}_{ps}^{\dagger} \hat{d}_{ps}^{} 
+ \hat{b}_{ps}^{\dagger} \hat{b}_{ps}^{} \Big) ~,
\end{eqnarray}
where all products of operators are taken in normal order.

From the Heisenberg equation of motion and the current
operator~(\ref{TotElCurr}) the force operator is found
\begin{eqnarray}
\label{JDotDir3}
\,\dot{\hat{\!j}}_{\mu}^{}(\vek{k}) &=& -\frac{ie}{\hbar}
\int \frac{d_{}^{3}p}{(2\pi\hbar)_{}^{3}}\;
\sqrt{\frac{m}{E_{p}^{}}} 
\sum_{ss'} \bigg\{
\nonumber\\[1pt]
& & ~~                                         
\sqrt{\frac{m}{E_{p+\hbar k}^{}}}
\hat{d}_{p,s}^{\dagger} \hat{d}_{p+\hbar k,s'}^{}\,
\Big( \bar{v}(p+\hbar k,s') \gamma_{\mu}^{} v(p,s) \Big) \,
\nonumber\\[1pt]
& & ~~
\Big[E_{p}^{}-E_{p+\hbar k}^{}\Big]\,
{\rm e}_{}^{\frac{i}{\hbar}(E_{p}^{}-E_{p+\hbar k}^{})t} \;
\nonumber\\[1pt]
&+&                                            
\sqrt{\frac{m}{E_{p-\hbar k}^{}}}
\hat{b}_{p-\hbar k,s'}^{\dagger} \hat{b}_{p,s}^{}\,
\Big( \bar{u}(p-\hbar k,s') \gamma_{\mu}^{} u(p,s) \Big) \,
\nonumber\\[1pt]
& & ~~
\Big[E_{p}^{}-E_{p-\hbar k}^{}\Big]\,
{\rm e}_{}^{-\frac{i}{\hbar}(E_{p}^{}-E_{p-\hbar k}^{})t} \;
\nonumber\\[1pt]
&+&                                            
\sqrt{\frac{m}{E_{p-\hbar k}^{}}}
\hat{d}_{-p+\hbar k,s'}^{} \hat{b}_{+p,s}^{}\,
\Big( \bar{v}(-p+\hbar k,s') \gamma_{\mu}^{} u(p,s) \Big) \,
\nonumber\\[1pt]
& & ~~
\Big[ E_{p}^{} + E_{p-\hbar k}^{}\Big]\,
{\rm e}_{}^{-\frac{i}{\hbar}(E_{p}^{}+E_{p-\hbar k}^{})t} \;
\nonumber\\[1pt]
&+&                                            
\sqrt{\frac{m}{E_{p+\hbar k}^{}}}
\hat{d}_{p,s}^{\dagger} \hat{b}_{-p-\hbar k,s'}^{\dagger}\,
\Big( \bar{u}(-p-\hbar k,s') \gamma_{\mu}^{} v(p,s) \Big) \,
\nonumber\\[1pt]
& & ~~
\Big[ E_{p}^{} + E_{p+\hbar k}^{}\Big]\,
{\rm e}_{}^{\frac{i}{\hbar}(E_{p}^{}+E_{p+\hbar k}^{})t} \;
\bigg\} ~.
\end{eqnarray}
Eq.~(\ref{TotElCurr}) and (\ref{JDotDir3}) can now be used 
to calculate the correlation function 
\begin{math}
\langle \,\hat{\!j}_{\!\mu}^{}(\vek{k})\, ;
\,\dot{\hat{\!j}}_{\!\nu}^{}(\vek{k}) \rangle_{\omega+i\eta}^{}
~.
\end{math}
The calculation is shown in Appendix~C. Together with the
Eqs.~(\ref{JJDot_Dir2}) -- (\ref{FermiRel:1}) as well as
Eq.~(\ref{Suscept1}) we have the 4-dimensional RPA susceptibility
tensor in the form
\begin{eqnarray}
\label{Chi_RPA1}
\lefteqn{
\chi_{\mu\nu}^{} (\vek{k},\omega)
=
\frac{e^2}{4} 
\int \frac{d_{}^{3}p}{(2\pi\hbar)_{}^{3}} \;
\frac{1}{E_{p}^{}} \; \bigg\{ }
\nonumber\\[1pt]
& & ~~                                           
\frac{-1}{E_{p+\hbar k}^{}} \,
\frac{\bar{f}(E_{p}^{}) - \bar{f}(E_{p+\hbar k}^{})}%
{\hbar\omega + E_{p}^{} - E_{p+\hbar k}^{} + i\eta} \;
{\rm tr}_{D}^{}\Big\{
\gamma_{\mu}^{} [\psl_{}^{} - m] \gamma_{\nu}^{} 
[\psl_{}^{}+\hbar \ksl - m] \Big\}
\nonumber\\
& & +                                            
\frac{1}{E_{p-\hbar k}^{}} \,
\frac{f(E_{p}^{}) - f(E_{p-\hbar k}^{})}%
{\hbar\omega - E_{p}^{} + E_{p-\hbar k}^{} + i\eta} \;
{\rm tr}_{D}^{}\Big\{
\gamma_{\mu}^{} [\psl_{}^{}+ m] \gamma_{\nu}^{} 
[\psl - \hbar \ksl + m] \Big\}
\nonumber\\
& & -                                            
\frac{1}{E_{p-\hbar k}^{}} \,
\frac{1- f(E_{p}^{}) - \bar{f}(E_{p-\hbar k}^{})}%
{\hbar\omega - E_{p}^{} - E_{p-\hbar k}^{} + i\eta} \;
{\rm tr}_{D}^{}\Big\{
\gamma_{\mu}^{} [\psl_{}^{} + m] \gamma_{\nu}^{} 
[-\psl + \hbar \ksl - m] \Big\}
\nonumber\\
& & +                                            
\frac{-1}{E_{p+\hbar k}^{}} \,
\frac{1- \bar{f}(E_{p}^{}) - f(E_{p+\hbar k}^{})}%
{\hbar\omega + E_{p}^{} + E_{p+\hbar k}^{} + i\eta} \;
{\rm tr}_{D}^{}\Big\{
\gamma_{\mu}^{} [\psl_{}^{} - m] \gamma_{\nu}^{} 
[-\psl - \hbar \ksl + m] \Big\} \bigg\} ~.
\end{eqnarray}
The fermion and anti-fermion distribution functions, $f$ and
$\bar{f}$ respectively, are defined by Eq.~(\ref{FFunc}).
The equation~(\ref{Chi_RPA1}) can be written in a more compact form,
if we notice, that the susceptibility tensor can be decomposed as
\begin{eqnarray}
\label{Chi4D_Decomp}
& &
\chi_{00}^{} = -\frac{k_{}^{i}}{k_{0}^{}}\chi_{i0}^{}
= -\frac{k_{}^{i}}{k_{0}^{}}\chi_{0i}^{}
\qquad , \qquad
\chi_{0i}^{} = \chi_{i0}^{} 
= -\frac{k_{i}^{}}{k_{0}^{}}\chi_{}^{\ell} ~,
\nonumber\\[4pt]
& &
\chi_{ij}^{} = \frac{k_{i}^{}k_{j}^{}}{\vek{k}_{}^{2}} \chi_{}^{\ell}
+ \bigg( \delta_{ij}^{} - \frac{k_{i}^{}k_{j}^{}}{\vek{k}_{}^{2}} \bigg)
\chi_{}^{t} 
\\[1pt] 
\label{Chi_L_T}
& &
\chi_{}^{\ell} = \frac{k_{i}^{}k_{j}^{}}{\vek{k}_{}^{2}} \chi_{ij}^{}
\qquad , \qquad
\chi_{}^{t} = \frac{1}{2} \Big[ \delta_{ij}^{} 
- \frac{k_{i}^{}k_{j}^{}}{\vek{k}_{}^{2}} \Big] \chi_{ij}^{} ~.
\end{eqnarray}
This means that the longitudinal and transverse component completely
determine the susceptibility tensor. This decomposition can be shown,
making use of the current conservation
\begin{math}
\omega \,\hat{\!j}_{0}^{} - \vek{k}\cdot \,\hat{\vek{\!j}} = 0~.
\end{math}

In order to give an expression for the $\chi_{}^{\ell}$ and
$\chi_{}^{t}$ we calculate the Dirac traces of the spatial part
$\chi_{}^{ij}$. For instance the first trace in Eq.~(\ref{Chi_RPA1})
yields 
\begin{eqnarray}
\label{DirTrace1}
& &
{\rm tr}_{D}^{} \Big\{
\gamma_{i}^{} [\psl - m] \gamma_{j}^{} [\psl + \hbar \ksl - m] \Big\}
\nonumber\\[1pt]
& &
= 4 \Big\{ 
\psl_{i}^{} (\psl + \hbar \ksl)_{j}^{}
+ (\psl + \hbar \ksl)_{i}^{} \psl_{j}^{}
+ \delta_{ij}^{} [E_{p}^{} E_{p+\hbar k}^{}
- \vek{p} (\vek{p}+\hbar\vek{k})
-m_{}^{2}]\Big\} ~.
\end{eqnarray}
Calculating the remaining traces accordingly and projecting the
longitudinal and transverse part [see Eq.~(\ref{Chi_L_T})]
as well as shifting $\vek{p}\to -\vek{p}$ in the first and fourth term
in Eq.~(\ref{Chi_RPA1}), we find
\begin{eqnarray}
\label{Chi_RPA2}
\lefteqn{
\chi_{}^{\ell,t} (\vek{k},\omega)
=
e^2
\int \frac{d_{}^{3}p}{(2\pi\hbar)_{}^{3}} \; \Bigg\{ }
\nonumber\\[2pt]
& & ~~
\Lambda_{-}^{\ell,t} \left[
\frac{f(E_{p}^{}) - f(E_{p-\hbar k}^{})}%
{\hbar\omega - E_{p}^{} + E_{p-\hbar k}^{} + i\eta} \,
- \frac{\bar{f}(E_{p}^{}) - \bar{f}(E_{p-\hbar k}^{})}%
{\hbar\omega + E_{p}^{} - E_{p-\hbar k}^{} + i\eta} 
\right]
\nonumber\\[2pt]
& & +                                            
\Lambda_{+}^{\ell,t} \left[
\frac{1- \bar{f}(E_{p}^{}) - f(E_{p-\hbar k}^{})}%
{\hbar\omega + E_{p}^{} + E_{p-\hbar k}^{} + i\eta} \,
- \frac{1- f(E_{p}^{}) - \bar{f}(E_{p-\hbar k}^{})}%
{\hbar\omega - E_{p}^{} - E_{p-\hbar k}^{} + i\eta} 
\right] \Bigg\} ~,
\end{eqnarray}
with the longitudinal and transverse projectors $\Lambda_{\pm}^{\ell,t}$
\begin{eqnarray}
\label{LamLong}
\Lambda_{\pm}^{\ell} &=& 1 \pm 
\frac{E_{p}^{2} + \hbar \vek{p}\vek{k} - 2
  \frac{(\vekind{p}\vekind{k})_{}^{2}}%
{\vekind{k}_{}^{2}}}{E_{p}^{} E_{p-\hbar k}^{}} ~,
\\[4pt]
\label{LamTrans}
\Lambda_{\pm}^{t} &=& 1 \pm 
\frac{m_{}^{2} - \hbar \vek{p}\vek{k} 
+ \frac{(\vekind{p}\vekind{k})_{}^{2}}%
{\vekind{k}_{}^{2}}}{E_{p}^{} E_{p-\hbar k}^{}} ~.
\end{eqnarray}
Observing, that $\Lambda_{\pm}^{\ell,t}$ is invariant under the shift
$\vek{p}\to -\vek{p}+\hbar \vek{k}$, we can perform this shift in all
terms containing $f(E_{p-\hbar k}^{})$ and $\bar{f}(E_{p-\hbar k}^{})$
in Eq.~(\ref{Chi_RPA2}). We finally have the result for the 
longitudinal and transverse RPA susceptibility tensor
\begin{eqnarray}
\label{Chi_RPA3}
\chi_{}^{\ell,t}(\vek{k},\omega) &=& e_{}^{2}
\int \frac{d_{}^{3}p}{(2\pi\hbar)_{}^{3}}\; F(E_{p}^{}) \Bigg\{
\frac{E_{p}^{}-E_{p-\hbar k}^{}}{(E_{p}^{}-E_{p-\hbar k}^{})^2 -
  \omega_{}^{2} - i\eta}
\Lambda_{-}^{\ell,t}
\nonumber \\[1pt]
& & +
\frac{E_{p}^{}+E_{p-\hbar k}^{}}{(E_{p}^{}+E_{p-\hbar k}^{})^2 -
  \omega_{}^{2} - i\eta }
\Lambda_{+}^{\ell,t} \Bigg\} 
+ \chi_{vac}^{\ell,t} ~,
\\[4pt]
\label{DistF}
F(E_{p}^{}) &=& 2\Big( f(E_{p}^{}) + \bar{f}(E_{p}^{}) \Big) ~,
\\[1pt]
\label{Chi_Vac}
\chi_{vac}^{\ell,t}(\vek{k},\omega)
&=& -2 e_{}^{2} \int \frac{d_{}^{3}p}{(2\pi\hbar)_{}^{3}}\;
\frac{E_{p}^{}+E_{p-\hbar k}^{}}{(E_{p}^{}+E_{p-\hbar k}^{})^2 -
  \omega_{}^{2} - i\eta }
\Lambda_{+}^{\ell,t} ~.
\end{eqnarray}
It should be noted that the result~(\ref{Chi_RPA3}) corresponds to
familiar expressions, for example published by
Tsytovich~\cite{Tsytovich1_61}, given in the form of the dielectric
tensor. 

The major difficulties in the theory arise, by the inclusion of
collisions, which will be the main issue of the next section. 

\subsection{Inclusion of collisions}
As in the case of the RPA approximation, we restrict ourselves to the
instant frame description, where the ions stay at rest 
($n_{\mu}^{}=(1,0,0,0)$). This also
implies that we only have to consider the Coulomb interaction term of
the electrons with the ions, which will be treated perturbatively.
In the case of an arbitrary reference frame, where the ions are
moving, we have 
\begin{math}
n_{}^{\mu}=(n_{}^{0},\vek{n})
=\left( \gamma , \vek{v}\gamma \right) ~,
\end{math}
where $\gamma$ is the relativistic factor 
$\gamma_{}^{-1} = \sqrt{1-\vek{v_{}^{2}}}$.

\begin{figure}[h]
\begin{minipage}{3.3cm}
\centerline{\psfig{figure=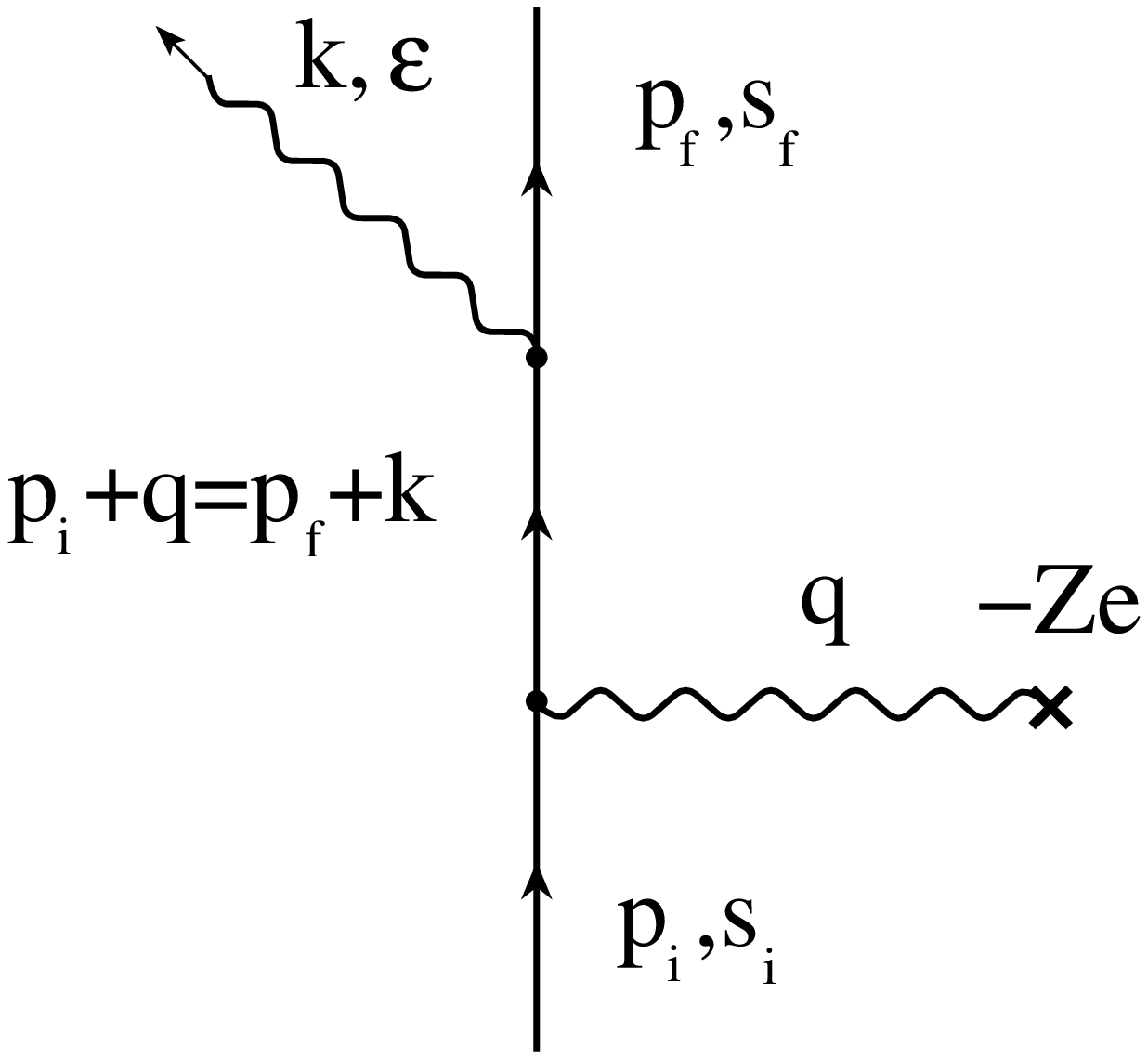,width=3cm}}
\end{minipage}
\begin{minipage}{3.3cm}
\centerline{\psfig{figure=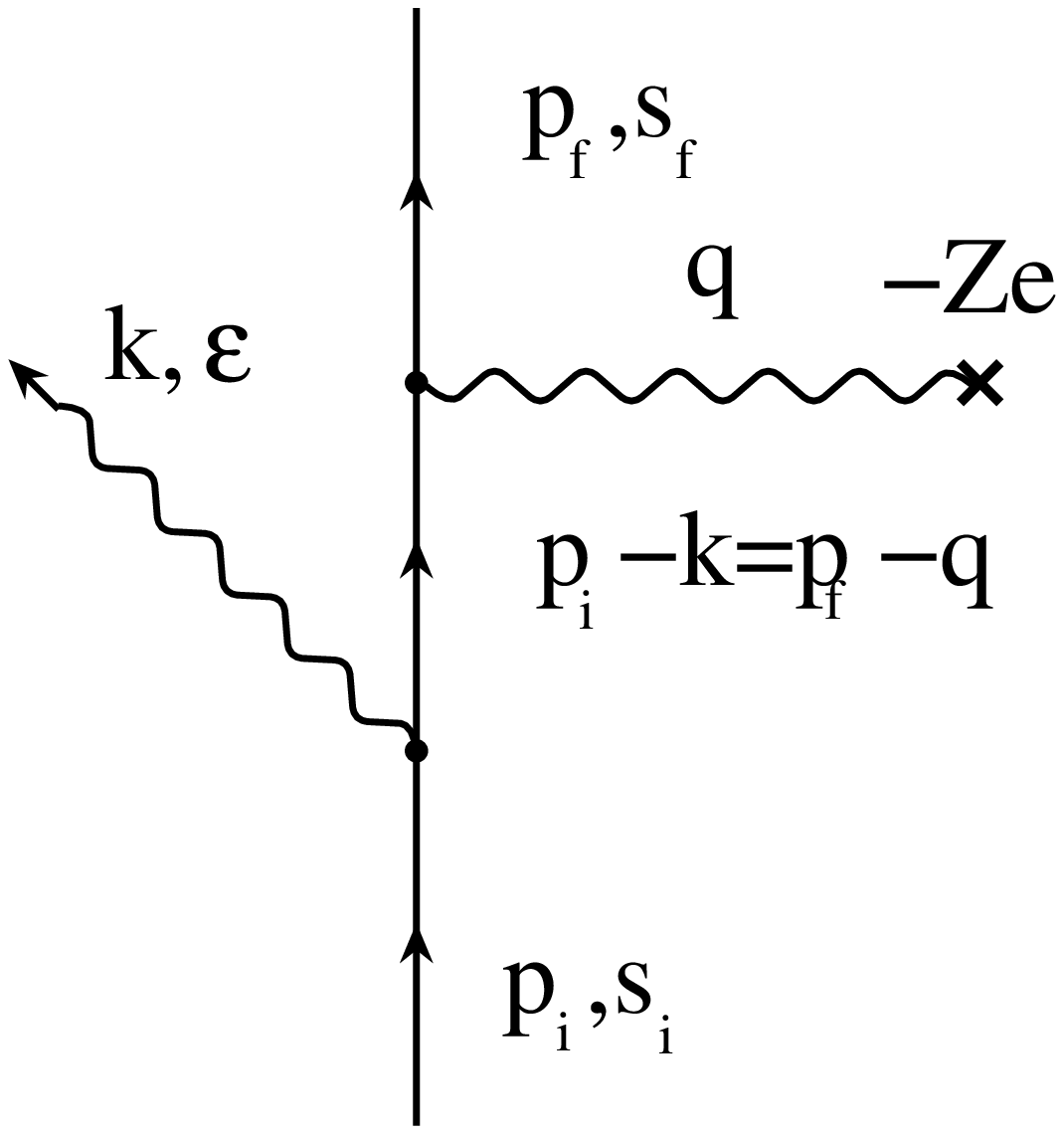,width=3cm}}
\end{minipage}
\hfill
\begin{minipage}{3.3cm}
\centerline{\psfig{figure=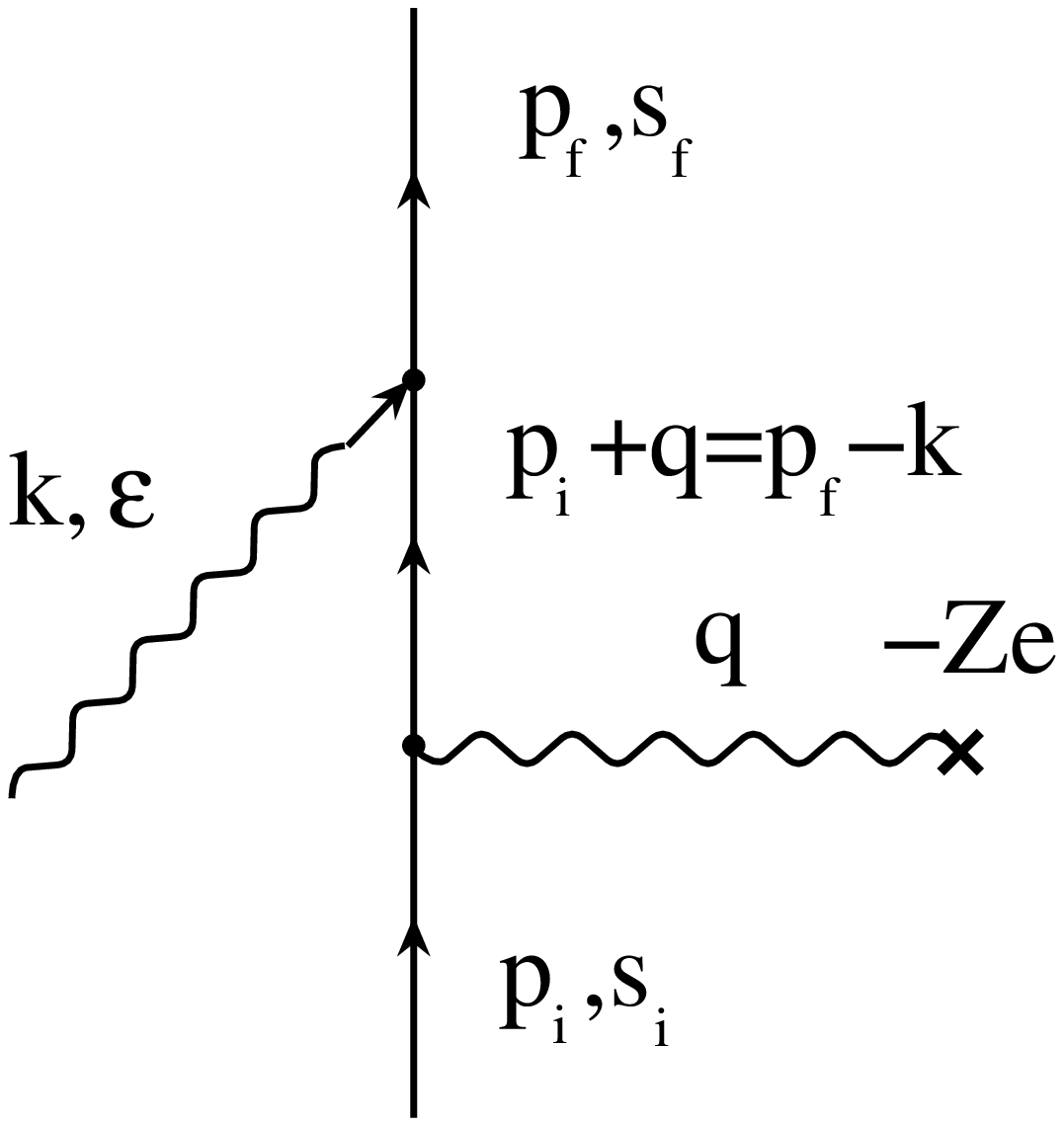,width=3cm}}
\end{minipage}
\begin{minipage}{3.3cm}
\centerline{\psfig{figure=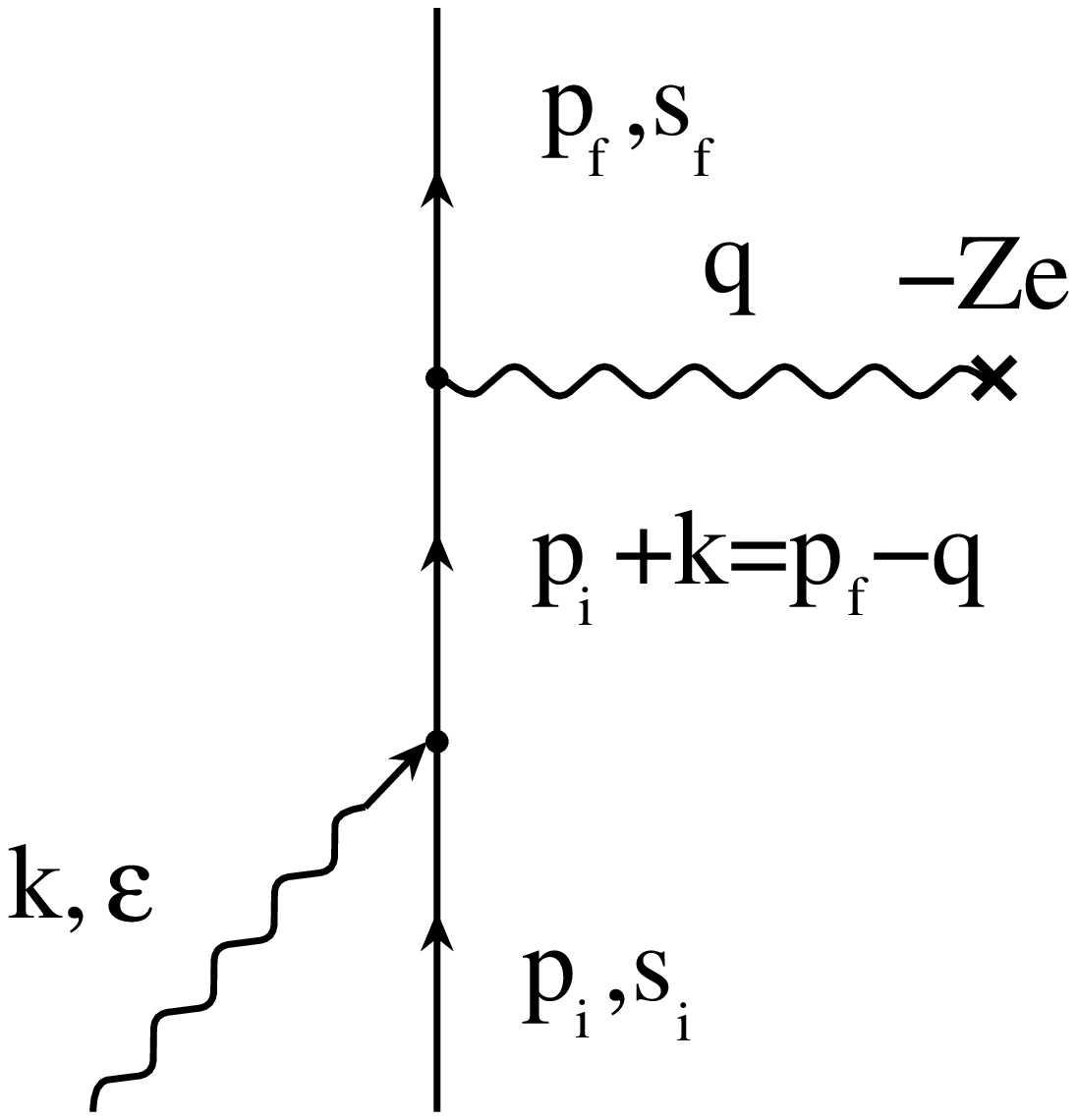,width=3cm}}
\end{minipage}
\\
\begin{minipage}[t]{6.5cm}
\caption{\label{P_FeynmEm} Feynman diagrams for second order
  bremsstrahlung of electrons in the Coulomb potential of the ions.}
\end{minipage}
\hfill
\begin{minipage}[t]{6.5cm}
\caption{\label{P_FeynmAb} Feynman diagrams for second order inverse
  bremsstrahlung of electrons in the Coulomb potential of the ions.}
\end{minipage}
\end{figure}

Collisions are included into the response function
Eq.~(\ref{Suscept1}) most conveniently, by making use of partial
integration and writing
$\chi_{\mu\nu}$ in terms of correlation functions including the force
operator $\,\dot{\hat{\!j}}_{\nu}^{}$ in the denominator. We can find
the matrix equation
\begin{eqnarray}
\label{Corr4D_01}
\Big\langle \,\hat{\!j}_{\mu}^{}\, ; \,\hat{\!j}_{\nu}^{} \Big\rangle
= - \frac{\left|
\begin{array}{cc}
0 & \Big( \,\hat{\!j}_{\mu}^{}\, ; \,\hat{\!j}_{\beta}^{} \Big) \\[1mm]
\Big( \,\hat{\!j}_{\alpha}^{}\, ; \,\hat{\!j}_{\nu}^{} \Big) & M_{\alpha\beta}^{}
\end{array} \right|}%
{\left| M_{\alpha\beta}^{} \right|} ~,
\end{eqnarray}
\begin{eqnarray}
\label{DefM_munu}
M_{\mu\nu}^{} 
= - i\omega \big( \,\hat{\!j}_{\mu}^{}\, ; \,\hat{\!j}_{\nu}^{} \big)
+ \big\langle \,\dot{\hat{\!j}}_{\mu}^{}\, ; \,\dot{\hat{\!j}}_{\nu}^{} \big\rangle
+ \frac{\left|
\begin{array}{cc}
0 & \langle \,\dot{\hat{\!j}}_{\mu}^{}\, ; \,\hat{\!j}_{\beta}^{} \rangle
\\[1mm]
\langle \,\hat{\!j}_{\alpha}^{}\, ;\,\dot{\hat{\!j}}_{\nu}^{} \rangle &
\langle \,\hat{\!j}_{\alpha}^{}\, ; \,\hat{\!j}_{\beta}^{} \rangle
\end{array}
\right|}%
{\left|\langle \,\hat{\!j}_{\alpha}^{}\, ; \,\hat{\!j}_{\beta}^{}
\rangle\right|} ~.
\end{eqnarray}
In Eq.~(\ref{DefM_munu}) the force-force correlation function appears,
which is well suited for a perturbative treatment.
Different processes can be found in each order of a perturbative
expansion. In QED they are well studied and can be represented by
Feynman diagrams.
In the Figs.~(\ref{P_FeynmEm}) and (\ref{P_FeynmAb}) we have shown the
second order diagrams for the bremsstrahlung and inverse bremsstrahlung
process. 
These processes are well known and expressions were given by QED
S-matrix calculations~\cite{Heitler} as well as by a non-relativistic
treatment~\cite{August}. We will draw our attention to the inverse
bremsstrahlung within our formalism in the next section.

\section{Inverse Bremsstrahlung}
As a second order process we consider in this section the
absorption of electro-magnetic waves in a relativistic plasma. For
illustration we derive the corresponding absorption coefficient in the
long wavelength limit.

\subsection{The absorption coefficient}
Before calculating the inverse bremsstrahlung in second order, we need
to give some relations, which allow to extract the absorption
coefficient $\alpha$ out of the 4-dimensional susceptibility tensor.

In Eq.~(\ref{Chi4D_Decomp}) and (\ref{Chi_L_T}) a general
decomposition of the susceptibility tensor into its longitudinal and
transverse part was shown.
The transverse component $\chi_{}^{t}$ can also be projected out of
$\chi_{\mu\nu}^{}$ by the two orthogonal transverse polarization vectors
$\epsilon_{\mu}^{}(k,1)$ or $\epsilon_{\mu}^{}(k,2)$
\begin{eqnarray}
\label{TransProChi}
\chi_{}^{t}(k) = \epsilon_{}^{\mu}(k,i)\, \chi_{\mu\nu}^{}(k)\,
\epsilon_{}^{\nu}(k,i) \quad , \quad i=1,2
\end{eqnarray}
While the susceptibility tensor describes the response of the plasma
due to an external field, the dielectric tensor
$\varepsilon_{\mu\nu}^{}$ connects the internal
and external fields with each other. The longitudinal and transverse
part of the dielectric tensor is related to the susceptibility
according to~\cite{Sitenko}
\begin{eqnarray}
\label{ChiL_EpsL}
\chi_{}^{\ell} (\vek{k},\omega) 
& = & 
\frac{\vek{k}^2}{e^2} \frac{ \varepsilon_{}^{\ell}(\vek{k},\omega)-1}{
\varepsilon_{}^{\ell}(\vek{k},\omega)} ~,
\\[3pt]
\label{ChiT_EpsT}
\chi_{}^{t}(\vek{k},\omega) 
&=& \frac{\vek{k}^2}{e^2} 
\left( 1-\frac{\vek{k}^2}{\omega^2} \right)\,
\frac{\varepsilon_{}^{t}(\vek{k},\omega)-1}{\varepsilon_{}^{t}(\vek{k},\omega)
-\vek{k}^2/\omega^2} ~.
\end{eqnarray}
In the long wavelength limit, which will be taken here, it is seen
from Eqs.~(\ref{ChiL_EpsL}) and (\ref{ChiT_EpsT}) that the
longitudinal and transverse part of the dielectric tensor and the
susceptibility tensor coincide. The long wavelength approximation is
well justified in the optical regime. 
The absorption coefficient is related to the damping of
electro-magnetic waves, which is determined by the imaginary part of the
dispersion equation. In the long wavelength limit we have 
\begin{eqnarray}
\label{AlphaEps:NR}
\alpha(\omega) = \frac{\omega}{n(\omega)} \lim_{\vekind{k}\to 0}\, 
{\rm Im}\;\varepsilon_{}^{t}(\omega,\vek{k}) ~,
\end{eqnarray}
where in the following the index of refraction is approximated by
$n(\omega)\approx 1$. 
From Eq.~(\ref{ChiT_EpsT}) we can find the relation
\begin{eqnarray}
\label{ImEpsT}
\lim_{\vekind{k}\to 0}{\rm Im}\, \varepsilon_{}^{t}(\omega) 
= \lim_{\vekind{k}\to 0} \frac{\vek{k}_{}^{2}}{e_{}^{2}} \;
\frac{{\rm Im\,\chi_{}^{t}}}{\frac{\vekind{k}^4}{e^4}
+ \Big[ {\rm Re\,\chi_{}^{t}} \Big]_{}^{2} 
+ \Big[ {\rm Im\,\chi_{}^{t}} \Big]_{}^{2}
- 2 \frac{\vekind{k}_{}^{2}}{e^2} {\rm Re\,\chi_{}^{t}} } ~,
\end{eqnarray}
with a $\vek{k}^4$-term dropped due to the limes $\vek{k}\to 0$.
It can be shown, that non-trivial solution for the transverse
susceptibility, making
use of Eq.~(\ref{Suscept1}) as well as Eqs.~(\ref{Corr4D_01}) and
(\ref{DefM_munu}), can be written as
\begin{eqnarray}
\label{ChiT}
\chi_{}^{t}(\vek{k},\omega) 
= - \beta \Big( \,\hat{\!j}_{\!\trans}^{} \,;\,\hat{\!j}_{\!\trans}^{}\Big)
- \frac{i \beta \omega \Big( \,\hat{\!j}_{\!\trans}^{}\,;
\,\hat{\!j}_{\!\trans}^{}\Big)
\Big( \,\hat{\!j}_{\!\trans}^{} \,;\,\hat{\!j}_{\!\trans}^{}\Big)}%
{-i\omega \left( \,\hat{\!j}_{\!\trans}^{} \,;
\,\hat{\!j}_{\!\trans}^{}\right)
+ \Big\langle \,\dot{\hat{\!j}}_{\trans}^{}\, ; 
\,\dot{\hat{\!j}}_{\trans}^{} \Big\rangle
} ~.
\end{eqnarray}
In these expressions we use the transverse current 
$\,\hat{\!j}_{\trans}^{}\equiv\epsilon_{}^{\mu}\,\hat{\!j}_{\mu}^{}$.
Contributions from the determinant, containing 
\begin{math}
\langle \,\hat{\!j}_{\alpha}^{}\, ;
\,\dot{\hat{\!j}}_{\nu}^{} \rangle
\end{math}
are neglected, since these terms are of higher order.
In order to derive Eq.~(\ref{ChiT}), the current conservation, leading
to the composition~(\ref{Chi4D_Decomp}) and (\ref{Chi_L_T}) was used.
Treating the collisions, i.e. 
\begin{math}
\Big\langle \,\dot{\hat{\!j}}_{\trans}^{}\, ; 
\,\dot{\hat{\!j}}_{\trans}^{} \Big\rangle
\end{math}
in Eq.~(\ref{ChiT})
as small contributions, we can write to first order in the force-force
correlation function
\begin{eqnarray}
\label{ChiT_Pert}
\chi_{}^{t}(\vek{k},\omega) \approx - \frac{i\beta}{\omega}  
\Big\langle \,\dot{\hat{\!j}}_{\trans}^{}\, ; 
\,\dot{\hat{\!j}}_{\trans}^{} \Big\rangle ~.
\end{eqnarray}
Finally with Eq.~(\ref{AlphaEps:NR}), (\ref{ImEpsT}) and
(\ref{ChiT_Pert}) we obtain the absorption coefficient containing
collisions in first order 
\begin{eqnarray}
\label{Alpha}
\alpha(\omega) = - \beta \lim_{\vekind{k}\to 0}
\frac{e_{}^{2}}{\vek{k}_{}^{2}}\,
\,{\rm Re}\;\Big\langle \,\dot{\hat{\!j}}_{\trans}^{}\, ; 
\,\dot{\hat{\!j}}_{\trans}^{} \Big\rangle_{\omega+i\eta}^{} ~.
\end{eqnarray}
The force-force correlation function in Eq.~(\ref{Alpha}) contains
both, the emission and the absorption of photons. The evaluation of
the correlation functions (see next section) shows that the absorption
is described by 
\begin{eqnarray}
\label{Alpha2}
\alpha(\omega) &=& -\beta \lim_{\vekind{k}\to 0}\,
\frac{e_{}^{2}}{\vek{k}_{}^{2}}\,
{\rm Re}\,\big\langle \,\dot{\hat{\!j}}_{\trans}^{}(-\vek{k})\, ; 
\,\dot{\hat{\!j}}_{\trans}^{}(-\vek{k}) \Big\rangle_{-\omega+i\eta}^{} ~,
\end{eqnarray}
with $\omega>0$, $\vek{k}>0$. 
The emission of photons, which will not be considered here further, is
related to the absorption by the interchange 
$\omega\to -\omega$ and $\vek{k}\to -\vek{k}$.

The relation, given in this section can straightforwardly generalized
to arbitrary hyperplanes. Since we work in the adiabatic approximation
all expressions were given in the instant frame, which is off course
most convenient here. 

\subsection{Evaluation of the correlation function}
Evaluating the force-force correlation function in Eq.~(\ref{Alpha2})
we will follow the same way presented in the last section.
The perturbative part in the Hamiltonian, i.e. the
Coulomb interaction with the ions, is written as
\begin{eqnarray}
\label{HIon}
\hat{H}_{ion}^{}
&=& e \int d_{}^{3}x \; \Norder\,
\hat{\!j}_{0}^{}(x)\Norder\, 
A_{ion}^{0}(\vek{x})
\nonumber\\[2pt]
&=& e \int 
d_{}^{3}p_{1}^{} \,
\frac{d_{}^{3}q}{(2\pi\hbar)_{}^{3}}\;
A_{ion}^{0}(\vek{q}) 
\sqrt{\frac{m}{E_{p_{1}^{}}}} 
\sum_{rr'} \Bigg\{ 
\nonumber\\[3pt]
& &                                             
\sqrt{\frac{m}{E_{p_1^{}-q}}} \,
\Norder 
\hat{d}_{p_{1}^{}-q,r'}^{} \hat{d}_{p_{1}^{},r}^{\dagger} \Norder \,
\Big( \bar{v}(p_1^{}-q,r') \gamma_{0}^{} v(p_{1}^{},r) \Big) \,
{\rm e}_{}^{\frac{i}{\hbar}(E_{p_{1}^{}}^{}-E_{p_{1}^{}-q}^{})t}
\nonumber\\[3pt]
&+& \sqrt{\frac{m}{E_{p_1+q}}} \,                     
\Norder 
\hat{b}_{p_{1}^{}+q,r'}^{\dagger} \hat{b}_{p_{1}^{},r}^{} \Norder \,
\Big( \bar{u}(p_{1}^{}+q,r') \gamma_{0}^{} u(p_{1}^{},r) \Big) \,
{\rm e}_{}^{-\frac{i}{\hbar}(E_{p_{1}^{}}^{}-E_{p_{1}^{}+q}^{})t}
\nonumber\\[3pt]
&+& \sqrt{\frac{m}{E_{-p_1^{}-q}}} \,         
\Norder 
\hat{d}_{-p_{1}^{}-q,r'}^{} \hat{b}_{p_{1}^{},r}^{} \Norder \,
\Big( \bar{v}(-p_{1}^{}-q,r') \gamma_{0}^{} u(p_{1}^{},r) \Big) \,
{\rm e}_{}^{-\frac{i}{\hbar}(E_{p_{1}^{}}^{}+E_{p_{1}^{}+q}^{})t}
\nonumber\\[3pt]
&+& \sqrt{\frac{m}{E_{-p_1^{}+q}}} \,         
\Norder 
\hat{b}_{-p_{1}^{}+q,r'}^{\dagger} \hat{d}_{p_{1}^{},r}^{\dagger} \Norder \,
\Big( \bar{u}(-p_{1}^{}+q,r') \gamma_{0}^{} v(p_{1}^{},r) \Big) \,
{\rm e}_{}^{\frac{i}{\hbar}(E_{p_{1}^{}}^{}+E_{p_{1}^{}-q}^{})t}
\Bigg\} ~.
\nonumber\\[3pt]
& & ~
\end{eqnarray}
In the ``rotating wave approximation'' (RWA) the last two terms in
Eq.~(\ref{HIon}) are neglected, since they describe rapid
oscillating processes with frequencies 
$\omega \approx 2m_{}^{}/\hbar$.  
We calculate 
\begin{math}
\,\dot{\hat{\!j}}_{\!\mu}^{}
=-\frac{i}{\hbar} [\,\hat{\!j}_{\!\mu}^{},\hat{H}_{ion}^{}]
\end{math}
using Eqs.~(\ref{TotElCurr}) and (\ref{HIon}) in the
RWA-approximation. 
After some algebra,
which is given in Appendix~D, we can find the expression
\begin{eqnarray}
\label{JDotRWA_Ion4}
\,\,\dot{\hat{\!\!j}}_{\mu}^{}(-\vek{k})
&=& -\frac{iem_{}^{}}{2\hbar} \int \frac{d_{}^{3}p_{i}^{}}{(2\pi\hbar)_{}^{3/2}} \,
\frac{d_{}^{3}p_{f}^{}}{(2\pi\hbar)_{}^{3/2}} \,
\sqrt{\frac{1}{E_{p_{i}^{}}^{}E_{p_{f}^{}}^{}}} \;
A_{ion}^{0}(\vek{q})
{\rm e}_{}^{\frac{i}{\hbar}
(E_{p_{f}^{}}^{}-E_{p_{i}^{}}^{})t} 
\nonumber\\[2pt]
& &
\sum_{s_{i}^{}s_{f}^{}}
\Big\{
N_{\mu}^{(p)} \;
\hat{d}_{p_{f}^{},s_{f}^{}}^{\dagger} \hat{d}_{p_{i}^{},s_{i}^{}}^{}
+ N_{\mu}^{(e)} \;
\hat{b}_{p_{f}^{},s_{f}^{}}^{\dagger} \hat{b}_{p_{i}^{},s_{i}^{}}^{}
\Big\} ~,
\\[4pt]
\label{DefN1}
& &
\hspace{-2cm}
N_{\mu}^{(p)} \;
=
\bar{v}(p_{i}^{},s_{i}^{}) \, {\rm tr}_D \bigg[ 
\gamma_{0}^{} \frac{\psl_{f}^{} - \hbar \ksl - m}{E_{p_{f}^{}-\hbar k}}
\gamma_{\mu}^{}
- \gamma_{\mu}^{}
\frac{\psl_{i}^{} + \hbar \ksl - m}{E_{p_{i}^{}+\hbar k}^{}}
\gamma_{0}^{} \bigg]
v(p_{f}^{},s_{f}^{}) \, ,
\\[2pt]
\label{DefN2}
& &
\hspace{-2cm}
N_{\mu}^{(e)} \;
=
\bar{u}(p_{f}^{},s_{f}^{}) \, {\rm tr}_D \bigg[ \gamma_{\mu}^{}
\frac{\psl_{f}^{} - \hbar \ksl + m}{E_{p_{f}^{}-\hbar k}}
\gamma_{0}^{}
- \gamma_{0}^{}
\frac{\psl_{i}^{} + \hbar \ksl + m}{E_{p_{i}^{}+\hbar k}}
\gamma_{\mu}^{} \bigg]
u(p_{i}^{},s_{i}^{}) \, .
\end{eqnarray}
It is now straightforward to calculate the force-force correlation
function, following the steps in the RPA-approxima\-tion (see
Appendix~C). As seen from Eq.~(\ref{Alpha2}) we need the expression
for the real part of the transverse force-force correlation.
With the definitions Eqs.~(\ref{Def:TCorrFns1}) and
(\ref{Def:LaplTCorrFns1}) we find after performing the contractions
using the Wick Theorem [Eq.~(\ref{WTheorem})] and integrating as well
as summing over the variables appearing in the delta-functions, we obtain
\begin{eqnarray}
\label{ReJDotJDot1}
& &
{\rm Re} 
\Big\langle
\,\dot{\hat{\!j}}_{\!\trans}^{} ,
\,\dot{\hat{\!j}}_{\!\trans}^{}
\Big\rangle_{-\omega+i\eta}^{-\vekind{k}}
= - \frac{e_{}^{2}m_{}^{2}}{4\hbar_{}^{2}\beta}\, {\rm Re} 
\int_{0}^{\infty} d\bar{t}\, 
{\rm e}_{}^{i(-\omega+i\eta)\bar{t}}
\int_{0}^{\beta} d \tilde{t}
\int \frac{d_{}^{3}p_{i}^{}}{(2\pi\hbar)_{}^{3}} \,
\frac{d_{}^{3}p_{f}^{}}{(2\pi\hbar)_{}^{3}} \;
\nonumber\\[2pt]
& &
\frac{1}{E_{p_{f}^{}}E_{p_{i}^{}}} \,
(A_{ion}^{0}(\vek{q}))_{}^{2}
{\rm e}_{}^{\frac{i}{\hbar}(E_{p_{f}^{}}^{} - E_{p_{i}^{}}^{})
(\bar{t}-i\hbar \tilde{t})}
\nonumber\\[2pt]
& &
\sum_{s_{i}^{}s_{f}^{}}
\bigg\{
\bar{f} (E_{p_{f}^{}}^{})\Big[1-\bar{f}(E_{p_{i}^{}}^{})\Big]
\left| \epsilon_{i}^{\mu} N_{\mu}^{(p)} \right|_{}^{2}
+ f (E_{p_{f}^{}}^{})\Big[1-f(E_{p_{i}^{}}^{})\Big]
\left| \epsilon_{i}^{\mu} N_{\mu}^{(e)} \right|_{}^{2}
\bigg\} ~.
\end{eqnarray}
The integration over $\bar{t}$ and $\tilde{t}$ can be performed, and
we obtain, with the Dirac identity 
\begin{math}
\lim_{\eta\to +0} (x\pm i \eta)_{}^{-1} = {\rm P} (1/x) \mp i\pi
\delta (x)
\end{math}
\begin{eqnarray}
\label{ReJDotJDot2}
& &
{\rm Re} 
\Big\langle
\,\dot{\hat{\!j}}_{\!\trans}^{} ,
\,\dot{\hat{\!j}}_{\!\trans}^{}
\Big\rangle_{-\omega+i\eta}^{}
= - \frac{e_{}^{2}m_{}^{2}\pi}{4\hbar_{}^{2}\omega\beta} 
\int \frac{d_{}^{3}p_{i}^{}}{(2\pi\hbar)_{}^{3}} \,
\frac{d_{}^{3}p_{f}^{}}{(2\pi\hbar)_{}^{3}} \;
\nonumber\\[2pt]
& &
\frac{\delta(E_{p_{f}^{}}^{} - E_{p_{i}^{}}^{} - \hbar\omega)}%
{E_{p_{f}^{}}E_{p_{i}^{}}} \;
(A_{ion}^{0}(\vek{q}))_{}^{2}
\Big( {\rm e}_{}^{\beta(E_{p_{f}^{}} - E_{p_{i}^{}})} - 1 \Big)
\nonumber\\[2pt]
& &
\sum_{s_{i}^{}s_{f}^{}}
\bigg\{
\bar{f} (E_{p_{f}^{}}^{})\Big[1-\bar{f}(E_{p_{i}^{}}^{})\Big]
\left| \epsilon_{i}^{\mu} N_{\mu}^{(p)} \right|_{}^{2}
+ f (E_{p_{f}^{}}^{})\Big[1-f(E_{p_{i}^{}}^{})\Big]
\left| \epsilon_{i}^{\mu} N_{\mu}^{(e)} \right|_{}^{2}
\bigg\} ~.
\end{eqnarray}
Using the first two relations of Eq.~(\ref{FermiRel:1}) as well as
Eq.~(\ref{Alpha2}) and writing
$E_{i} \equiv E_{p_{i}^{}}$ and $E_{f}^{} \equiv E_{p_{f}^{}}$ 
\begin{eqnarray}
\label{ReJDotJDot3}
& &
{\rm Re} 
\Big\langle
\,\dot{\hat{\!j}}_{\!\trans}^{} ,
\,\dot{\hat{\!j}}_{\!\trans}^{}
\Big\rangle_{-\omega+i\eta}^{}
= \frac{e_{}^{2}m_{}^{2}\pi}{4\hbar_{}^{2}\omega\beta} 
\int \frac{d_{}^{3}p_{i}^{}}{(2\pi\hbar)_{}^{3}} \,
\frac{d_{}^{3}p_{f}^{}}{(2\pi\hbar)_{}^{3}} \;
\frac{\delta(E_{f} - E_{i} - \hbar\omega)}%
{E_{p_{f}^{}}E_{p_{i}^{}}} \;
(A_{ion}^{0}(\vek{q}))_{}^{2}
\nonumber\\[2pt]
& &
\sum_{s_{i}^{}s_{f}^{}}
\bigg\{
\left| \epsilon_{i}^{\mu} N_{\mu}^{(p)} \right|_{}^{2}
\Big[ \bar{f}(E_{f}^{}) - \bar{f}(E_{i}^{}) \Big]
+ 
\left| \epsilon_{i}^{\mu} N_{\mu}^{(e)} \right|_{}^{2}
\Big[ f(E_{f}^{}) - f(E_{i}^{}) \Big]
\bigg\} ~.
\end{eqnarray}
Finally with Eq.~(\ref{Alpha2}) we find for the absorption coefficient
\begin{eqnarray}
\label{Alpha3}
& &
\alpha(\omega)
= \lim_{\vekind{k}\to 0}\,
\frac{e_{}^{4}m_{}^{2}\pi}{4\hbar_{}^{2}\omega_{}^{}\vek{k}_{}^{2}} 
\int \frac{d_{}^{3}p_{i}^{}}{(2\pi\hbar)_{}^{3}} \,
\frac{d_{}^{3}p_{f}^{}}{(2\pi\hbar)_{}^{3}} \;
\frac{\delta(E_{f} - E_{i} - \hbar\omega)}%
{E_{p_{f}^{}}E_{p_{i}^{}}} \;
(A_{ion}^{0}(\vek{q}))_{}^{2}
\nonumber\\[2pt]
& &
\sum_{s_{i}^{}s_{f}^{}}
\bigg\{
\left| \epsilon_{i}^{\mu} N_{\mu}^{(p)} \right|_{}^{2}
\Big[ \bar{f}(E_{f}^{}) - \bar{f}(E_{i}^{}) \Big]
+ 
\left| \epsilon_{i}^{\mu} N_{\mu}^{(e)} \right|_{}^{2}
\Big[ f(E_{f}^{}) - f(E_{i}^{}) \Big]
\bigg\} ~.
\end{eqnarray}
It should be emphazised, that the transition matrices $N_{\mu}^{(p)}$
and $N_{\mu}^{(e)}$ carry a $\vek{k}-$dependence (see
Eqs.~(\ref{DefN1}) and (\ref{DefN2})) leading to a finite
absorption coefficient in Eq.~(\ref{Alpha3}). As already pointed out
the polarization vector $\epsilon_{i}^{\mu}$ for a linear polarized
wave is one of the two transverse modes, labeled here by $i=1,2$.

Within the approximation applied here, we observe from
Eq.~(\ref{Alpha3}), that we have an electron as well as a positron
contribution, responsible for the absorption of the external wave. 
In first order perturbation theory these two terms are
not coupled to each other. 
Writing the result in terms of the transition matrices $N_{\mu}^{(p)}$
and $N_{\mu}^{(e)}$, it can further be observed, that our result
corresponds to the well known Bethe-Heitler~\cite{Heitler} formula, if
treating the absorption of the wave by a single incoming electron with energy
$E_i$ and an outgoing electron of energy $E_f$. 

In~\cite{August} non-relativistic results for the absorption
coefficient in different approximations are derived. The expression
given in the paper in terms of the transition matrix correspond to the
non-relativistic result of Eq.~(\ref{Alpha3}) if the matrix element is
evaluated in Born approximation. The final result is written for the
complex collision frequency $\nu$, which is related to the absorption
coefficient $\alpha$ by
\begin{math}
\label{Alpha_Nu}
\alpha(\omega) = \omega_{pl}^{2}/\omega_{}^{2}\,{\rm Re}\,\nu(\omega) ~,
\end{math}
with $\omega_{pl}^{}$ the electron plasma frequency and the index of
refraction $n(\omega)=1$.

\section{Discussion and Outlook}
We demonstrated in this work, how the hyperplane formalism can be
used for a manifest covariant density matrix formulation of
relativistic plasmas. The one-time description allows to formulate an
initial value problem, which can lead to considerable simplifications in
short-time pump-and probe experiments.
A covariant scheme is developed in the hyperplane formalism, where
Heisenberg operators are defined on spacelike hyperplanes in Minkowski
space.
In particular, the construction of the quantum Hamiltonian starting
from the classical QED Lagrangian, making use of the canonical
quantization scheme is shown.

From the Liouville von Neumann equation an initial value problem in
the hyperplane formalism is formulated. The approach used in this work is a
generalization of Zubarev's method of the relevant statistical
operator. For the case of moderate fields we applied the 
linear response approximation, where the statistical operator is expanded
near its equilibrium solution. 
From the self-consistency relation, which determines the Lagrange
multipliers of the generalized Gibbsian ensemble, we obtain 
the response equation, defining the susceptibility tensor on the
hyperplane in terms of current-current correlation functions.
The relativistic susceptibility tensor, which displays the response of
the fermion current to an external electro-magnetic wave, is
calculated in the RPA approximation. The result agrees with familiar
results, published for instance by Tsytovich~\cite{Tsytovich1_61}.

Further it was demonstrated how to include collisions into the
formalism in a systematic way, making use of perturbation theory. 
For that reason the current-current correlation function is expressed
by force-force correlation functions by partial integration. The
force-force correlation functions contain collisions since the force
operators are calculated from the von Neumann equation with the
interaction part in the Hamiltonian.
The advantage of the representation of the current-current correlation
function, is the appearance of the force-force correlation function in
denominator, which is convenient for a perturbative expansion. 

As an illustration, we derived an expression for the absorption
coefficient of inverse bremsstrahlung in first order of the
force-force correlation function (which corresponds to second order in
the interaction). We made use of the adiabatic approximation, where
the dynamic of the positively charged plasma component is frozen as
well as the Born approximation in calculating the correlation functions. 
The result given here is a generalization of the Bethe-Heitler formula
for the case the absorption in a electron-positron plasma. The
interaction is described by the Coulomb interaction.

Extensions to the approximations assumed here in this work can be done
in different ways. It is possible to consider higher orders in the
expansion of the current-current correlation function in terms of the
force-force correlation function. This will allow to describe higher
order processes, like for instance pair production.
Further interactions can be taken into account, like the
electron-electron interaction, or the radiation part of
the Hamiltonian, which implies to couple photons into the plasma.

A general scheme to derive kinetic equations in the hyperplane formalism is
explained in~\cite{hoell1}, valid also for the case of strong external
fields.
However, processes beyond the RPA approximation are hard to be
calculated, since coupled kinetic equations for the fermions and
photons are to be solved.

\renewcommand{\theequation}{A.\arabic{equation}}
\setcounter{equation}{0}
\section*{Appendix A}
\subsection*{Commutation relations on hyperplanes}
The constraint equations for the canonical variables
$A^{\mu}_{\trans}$ and $\Pi^{\mu}_{\trans}$  on the hyperplane
$\sigma_{n,\tau}^{}$ is written in the form
$\chi^{}_{N}(x^{}_{\trans})=0$, where
\begin{equation}
\label{ConstrFunc}
\begin{array}{ll}
\chi^{}_{1}(x^{}_{\trans})=
\nabla^{}_{\mu} A^{\mu}_{\trans}(x^{}_{\trans}),
&
\qquad
\chi^{}_{2}(x^{}_{\trans})=
\nabla^{}_{\mu} \Pi^{\mu}_{\trans}(x^{}_{\trans}),
\\[6pt]
\chi^{}_{3}(x^{}_{\trans})=
n^{}_{\mu} A^{\mu}_{\trans}(x^{}_{\trans}),
&
\qquad
\chi^{}_{4}(x^{}_{\trans})=
n^{}_{\mu}  {\Pi}^{\mu}_{\trans}(x^{}_{\trans}).
\end{array}
\end{equation}
For any functionals $\Phi^{}_{1}$ and $\Phi^{}_{2}$
of the field variables $A^{}_{\trans}$ and $\Pi^{}_{\trans}$,
we define the Poisson bracket
\begin{equation}
\label{P-brack}
\hspace*{-15pt}
\left[\Phi^{}_{1},\Phi^{}_{2}\right]^{}_{\rm P}\equiv
\int^{}_{\sigma^{}_{n,\tau}} d\sigma
\left\{
\frac{\delta\Phi^{}_{1}}
{\delta A^{\mu}_{\trans}(x^{}_{\trans})}\,
\frac{\delta\Phi^{}_{2}}
{\delta \Pi^{}_{\trans\mu}(x^{}_{\trans})}
-\frac{\delta\Phi^{}_{2}}
{\delta A^{\mu}_{\trans}(x^{}_{\trans})}\,
\frac{\delta\Phi^{}_{1}}
{\delta \Pi^{}_{\trans\mu}(x^{}_{\trans})}
\right\},
\end{equation}
where the constraints are ignored in calculating the functional
derivatives. Applying this formula to the canonical variables we obtain
\begin{equation}
\label{APi:P-br}
\left[
A^{\mu}_{\trans}(x^{}_{\trans}),
\Pi^{}_{\trans\nu}(x^{\prime}_{\trans})
\right]^{}_{\rm P}=\delta^{\mu}_{\ \nu}\,
\delta^{3}(x^{}_{\trans}-x^{\prime}_{\trans})
\end{equation}
with the three-dimensional delta function~(\ref{DeltaFunc}).
All other Poisson brackets for the canonical variables are equal to
zero. In the Dirac terminology,  functions~(\ref{ConstrFunc})
correspond to {\em second class\/} constraints since the matrix
\begin{equation}
\label{Cmatr}
C^{}_{NN'}(x^{}_{\trans}, x^{\prime}_{\trans})=
\left[
\chi^{}_{N}(x^{}_{\trans}),\chi^{}_{N'}(x^{\prime}_{\trans})
\right]^{}_{\rm P}
\end{equation}
is non-singular. A straightforward calculation of the Poisson brackets
shows that the non-zero elements of $C$ are
\begin{eqnarray}
\label{Cmatr:el}
& &
C^{}_{12}(x^{}_{\trans}, x^{\prime}_{\trans})=
- C^{}_{21}(x^{}_{\trans}, x^{\prime}_{\trans})=
-\nabla^{}_{\mu} \nabla^{\mu}
\delta^3(x^{}_{\trans}-x^{\prime}_{\trans}),
\nonumber\\[6pt]
& &
C^{}_{34}(x^{}_{\trans}, x^{\prime}_{\trans})=
- C^{}_{43}(x^{}_{\trans}, x^{\prime}_{\trans})=
\delta^3(x^{}_{\trans}-x^{\prime}_{\trans}).
\end{eqnarray}
According to the general quantization scheme~\cite{Dirac50,Weinberg96},
commutation relations for canonical operators
are defined by the Dirac brackets for classical canonical
variables. In our case  the Dirac brackets are written as
\begin{eqnarray}
\label{DiracBr:Def}
& &
\hspace*{-10pt}
\left[\Phi^{}_{1},\Phi^{}_{2}\right]^{}_{\rm D}=
\left[\Phi^{}_{1},\Phi^{}_{2}\right]^{}_{\rm P}
\nonumber\\[8pt]
& &
\hspace*{30pt}
{}- \int_{\sigma^{}_{n,\tau}} d\sigma
\int_{\sigma^{}_{n,\tau}} d\sigma'
\left[\Phi^{}_{1},\chi^{}_{N}(x^{}_{\trans})\right]^{}_{\rm P}
C^{-1}_{NN'}(x^{}_{\trans},x^{\prime}_{\trans})
\left[\chi^{}_{N'}(x^{\prime}_{\trans}),
\Phi^{}_{2}\right]^{}_{\rm P}
\end{eqnarray}
(summation over repeated indices). The inverse matrix,
$C^{-1}_{NN'}(x^{}_{\trans},x^{\prime}_{\trans})$, satisfies the
equation
\begin{equation}
\label{InverC:Eq}
\int_{\sigma^{}_{n,\tau}} d\sigma''\,
C^{}_{NN''}(x^{}_{\trans},x^{\prime\prime}_{\trans})\,
C^{-1}_{N''N'}(x^{\prime\prime}_{\trans},x^{\prime}_{\trans})=
\delta^{}_{NN'}\,\delta^{3}(x^{}_{\trans}-x^{\prime}_{\trans}).
\end{equation}
Since the matrix elements~(\ref{Cmatr:el}) of $C$ depend on the
difference $x^{}_{\trans}-x^{\prime}_{\trans}$, Eq.~(\ref{InverC:Eq})
can  be solved for $C^{-1}$ using a Fourier transform on
$\sigma^{}_{n,\tau}$, which is defined for any function $f(x)$ as
\begin{equation}
\label{FourTrans}
\tilde{f}(\tau,p^{}_{\trans})=
\int d^4x\, {\rm e}^{ip_{\!\trans}^{}\cdot x_{\!\trans}^{}}\, 
\delta(x\cdot n -\tau)\,f(x).
\end{equation}
The inverse transform is
\begin{equation}
\label{FourTrans:Inv}
f(x)\equiv f(\tau,x^{}_{\trans})=
\int \frac{d^{4}p}{(2\pi)^{3}}\,
{\rm e}^{-ip\cdot x}\, \delta(p\cdot n) \tilde f(\tau,p^{}_{\trans}).
\end{equation}
If we perform the Fourier transformation in Eq.~(\ref{InverC:Eq}), we
find by inserting~(\ref{Cmatr:el}) that  the non-zero
elements of $C^{-1}$ are
\begin{eqnarray}
\label{C-1:el}
& &
C^{-1}_{12}(x^{}_{\trans},x^{\prime}_{\trans})=
- C^{-1}_{21}(x^{}_{\trans},x^{\prime}_{\trans})=
- \int \frac{d^{4}p}{(2\pi)^{3}}\,
{\rm e}^{-ip\cdot (x -x')}\, \delta(p\cdot n) \frac{1}{p^{2}_{\trans}},
\nonumber\\[6pt]
& & C^{-1}_{34}(x^{}_{\trans},x^{\prime}_{\trans})= -
C^{-1}_{43}(x^{}_{\trans},x^{\prime}_{\trans})= -
\delta^{3}(x^{}_{\trans}-x^{\prime}_{\trans}).
\end{eqnarray}
Now the Dirac brackets~(\ref{DiracBr:Def}) for the canonical variables
are easily calculated and we obtain
\begin{eqnarray}
\label{DiracBr:Can1}
& & \left[
A^{\mu}_{\trans}(x^{}_{\trans}), {\Pi}^{\nu}_{\trans}(x^{\prime}_{\trans})
\right]^{}_{\rm D}= c^{\mu\nu}(x^{}_{\trans} -x^{\prime}_{\trans}),
\\[6pt]
& &
\label{DiracBr:Can2}
\left[A^{\mu}_{\trans}(x^{}_{\trans}),
{A}^{\nu}_{\trans}(x^{\prime}_{\trans})
\right]^{}_{\rm D}=\left[{\Pi}^{\mu}_{\trans}(x^{}_{\trans}),
{\Pi}^{\nu}_{\trans}(x^{\prime}_{\trans})
\right]^{}_{\rm D}=0,
\end{eqnarray}
where the functions $c^{\mu\nu}(x^{}_{\trans} -x^{\prime}_{\trans})$ are given
by Eq.~(\ref{DiracDelta}).
According to the general quantization rules, the commutation relations
for canonical operators correspond to $i[\ldots]^{}_{\rm D}$.
Thus, in the hyperplane formalism, the commutation relations for the
operators of EM field are given by~(\ref{Comm:Can}) and~(\ref{Comm:CanZero}).
Obviously these relations are valid in the
Schr\"odinger and Heisenberg pictures.

\subsection*{The anti-commutation relations on hyperplanes}
To find the anticommutation relations for the fermion
operators  on the hyperplane $\sigma^{}_{n,\tau}$, it is sufficient
to consider a free Dirac field. Our starting point is the standard
quantization scheme in the frame where
$x^{\mu}=(t,\vek{r})$ and $n^{\mu}=(1,0,0,0)$
(see, e.g.,~\cite{Gross}).
In that case the field  operators $\hat{\psi}^{}_{a}$ and
$\,\hat{\!\bar\psi}^{}_{a}$
can be written in terms of creation and
annihilation operators according to
$$
\begin{array}{l}
\displaystyle
\hat{\psi}_{a}^{}(x) = \int \frac{d^4 p}{(2\pi )_{}^{3/2}} \;
\frac{\delta (p_{}^{0}
- \epsilon (\vek{p}))}{\sqrt{2 \epsilon (\vek{p})}} \sum_{s=\pm 1}
\left[
\hat{b}_{s}^{}(p) u_{as}^{}(p) \e_{}^{-ip\cdot x}
+ \hat{d}_{s}^{\dagger}(p) v_{as}^{}(p)\e_{}^{ip\cdot x}
\right],
\\[18pt]
\displaystyle
\,\hat{\!\bar\psi}_{\!a}^{}(x) = \int \frac{d^4 p}{(2\pi )_{}^{3/2}} \;
\frac{\delta (p_{}^{0}
- \epsilon (\vek{p}))}{\sqrt{2 \epsilon (\vek{p})}} \sum_{s=\pm 1}
\left[
\hat{d}_{s}^{}(p) \bar v_{as}^{}(p) \e_{}^{-ip\cdot x}
+ \hat{b}_{s}^{\dagger}(p) \bar u_{as}^{}(p) \e_{}^{ip\cdot x}
\right],
\end{array}
$$
where $\epsilon (\vek{p}) =\sqrt{\vek{p}_{}^{2} + m_{}^{2}}$
is the free fermion dispersion relation.
Constructing the
expression
$\{\hat{\psi}_{a}^{}(x),\,\hat{\!\bar\psi}_{\!a'}^{}(x')\}$
for two arbitrary space-time points and recalling
the anticommutation relations
\begin{equation}
\label{Acommbd}
\left\{
\hat{b}_{s}^{}(\vek{p}) , \hat{b}_{s'}^{\dagger}(\vek{p}')
\right\}
=
\left\{
\hat{d}_{s}^{}(\vek{p}) , \hat{d}_{s'}^{\dagger}(\vek{p}')
\right\}
= \delta_{ss'} \delta_{}^{3}(\vek{p} - \vek{p}'),
\end{equation}
as well as polarization sums
$$
\sum_{s=\pm 1} u^{}_{as}(p)\bar{u}^{}_{a's}(p)=
\left[\gamma^{\mu}p^{}_{\mu} +m \right]^{}_{aa'},
\qquad
\sum_{s=\pm 1} v^{}_{as}(p)\bar{v}^{}_{a's}(p)=
\left[\gamma^{\mu}p^{}_{\mu} - m \right]^{}_{aa'},
$$
we arrive at
\begin{eqnarray}
\label{Tmp1}
\left\{
\hat{\psi}_{a}^{}(x),\,\hat{\!\bar\psi}_{\!a'}^{}(x')
\right\}
&=& \int \frac{d^3\vek{p}}{(2\pi )_{}^{3}} \;
\frac{1}{2\epsilon (\vek{p})}
\left\{
\left[\gamma^{\mu} p^{}_{\mu}
 +  m \right]_{aa'}
\e^{-ip\cdot (x-x')}
\right.
\nonumber\\[6pt]
& &
\hspace*{60pt}
\left.
+ \left[ \gamma^\mu p^{}_\mu -  m \right]_{aa'}
\e^{ip\cdot (x-x')} \right\},
\end{eqnarray}
where $p^{0} = \sqrt{\vek{p}_{}^{2} + m^2}$.
Using
\begin{equation}
\label{Tmp3}
\int \frac{d^3 \vek{p}}{(2 \pi )^3} \;
\frac{1}{2 \epsilon (\vek{p})}
=
\left.
\int \frac{d^4 p}{(2 \pi )^3}\;
\delta (p^2 - m^2) \right|_{p^{0} > 0}\, ,
\end{equation}
Eq.~(\ref{Tmp1}) can be rewritten in a Lorentz invariant form
\begin{eqnarray}
\label{Tmp4}
& &
\hspace*{-25pt}
\left\{
\hat{\psi}_{a}^{}(x),\,\hat{\!\bar\psi}_{a'}^{}(x')
\right\}
=
\int \frac{d^4 p}{(2 \pi )^3}\;
\left\{
\left[\gamma^\mu  p^{}_\mu +  m \right]_{aa'}
\e^{-ip\cdot (x-x')}
\left. \delta (p^2 - m^2)\right|_{p^0 > 0}
\right.
\nonumber\\[6pt]
& &
\hspace*{80pt}
\left.
+ \left[ \gamma^\mu  p^{}_\mu -  m \right]_{aa'}
\e^{ip\cdot (x-x')}
\left. \delta (p^2 - m^2)\right|_{p^0 > 0}
\right\}.
\end{eqnarray}
The  anticommutation relation on the hyperplane $\sigma^{}_{n,\tau}$
is now obtained by setting $x=n\tau +x^{}_{\trans}$ and
$x'=n\tau +x^{\prime}_{\trans}$. In calculating the integrals, it is
convenient to use the decomposition
$p^{\mu}= n^{\mu} p^{}_{\longi} + p^{\mu}_{\trans}$,
($p^{}_{\longi}>0$).
Then we get
\begin{eqnarray}
\label{Tmp5}
& &
\hspace*{-20pt}
\left\{
\hat{\psi}_{a}^{}(\tau ,x_{\trans}^{}),
\,\hat{\!\bar\psi}_{\!a'}^{}(\tau ,x_{\trans}^{\prime} )
\right\}
= \int \frac{d^4 p}{(2\pi )^3}\,
\frac{\delta (p_{\longi}^{ } - \epsilon (p_{\trans}^{}))}
{2\epsilon (p_{\trans}^{})}
\nonumber\\[8pt]
& &
\hspace*{60pt}
{}\times\left\{
\left[
\gamma_{\longi}^{} p_{\longi}^{}
+ \gamma^{\mu}_{\trans} p^{}_{\trans\mu} +  m
\right]_{aa'}^{}
\e^{-ip_{\trans\mu}^{} (x_{\trans}^{\mu} - x_{\trans}^{\prime\mu})}
\right.
\nonumber\\[6pt]
& &
\hspace*{120pt}
\left.
{}+
\left[
\gamma_{\longi}^{}p_{\longi}^{}
+ \gamma^{\mu}_{\trans}  p^{}_{\trans\mu} -  m
\right]_{aa'}^{}
\e^{ip_{\trans \mu}^{}(x_{\trans}^{\mu} - x_{\trans}^{\prime\mu})}
\right\}
\end{eqnarray}
with the dispersion relation on the hyperplane
\begin{equation}
\label{Disp}
\epsilon (p_{\trans}^{})
= \sqrt{-p_{\trans \mu}^{} p_{\trans}^{\mu} + m^{2}}.
\end{equation}
Finally, changing  the variable
$p_{\trans} \rightarrow - p_{\trans}$ in the second integral in
Eq.~(\ref{Tmp5}), we obtain the anticommutation
relation~(\ref{Anticomm1}). The relations~(\ref{Anticomm2}) can be derived
by the same procedure.

\renewcommand{\theequation}{B.\arabic{equation}}
\setcounter{equation}{0}
\section*{Appendix B}

\subsection*{The relevant statistical operator in linear response}
In order to rewrite Eq.~(\ref{Rho:Relevant}), we make use of the
operator identity
\begin{eqnarray}
\label{Op:Expansion}
{\rm e}_{}^{\hat{C}_{1} + \hat{C}_{2}^{}}
= \left(
1 + \int_{0}^{1}dz \; {\rm e}_{}^{z(\hat{C}_{1}^{} + \hat{C}_{2}^{})}
\,\hat{C}_{2}^{}\, {\rm e}_{}^{-z\hat{C}_{1}^{}} 
\right) {\rm e}_{}^{\hat{C}_{1}^{}}
\end{eqnarray}
and obtain
\begin{eqnarray}
\label{Rho:RelLin}
\varrho_{{\rm rel}}^{}(n,\tau) 
&=& \Bigg(
1 + \beta \int_{0}^{1} dz \,\bigg\{
{\rm e}_{}^{-z \beta \big[ \hat{H}_{s}^{} 
- \nu_{}^{} \hat{Q}_{}^{} \big]}
\nonumber\\[4pt]
& &
\times \int_{\sigma_{n}^{}} d\sigma ~ \sum_{\ell}\phi_{\mu}^{\ell}(x_{\trans}^{};\tau) 
\hat{B}_{\ell}^{\mu}(x_{\trans}^{})
{\rm e}_{}^{z \beta \big[ \hat{H}_{s}^{} 
- \nu_{}^{} \hat{Q}_{}^{} \big]}
\bigg\}\Bigg) \varrho_{0}^{}(n) ~.
\end{eqnarray}
The Lagrange multipliers $\phi_{\mu}^{\ell}$ are already first order
contributions for the deviation from the equilibrium. Therefore in
zeroth order, we can express the Schr\"odinger operators 
$\hat{B}_{\ell}^{\mu}(x_{\trans}^{})$ by some $\tau$-dependent
Heisenberg operators according to
 \begin{eqnarray}
\label{B:Heisenberg}
{\rm e}_{}^{-z \beta \hat{H}_{s}^{}}
\,\hat{B}_{\ell}^{\mu}(x_{\trans}^{},\tau_{0}^{})\,
{\rm e}_{}^{z \beta \hat{H}_{s}^{}} 
&=& U_{0}^{\dagger}(i z \beta,\tau_{0}^{})
\,\hat{B}_{\ell}^{\mu}(x_{\trans}^{},\tau_{0}^{})\,
U_{0}^{}(i z \beta,\tau_{0}^{})
\nonumber\\[1mm]
&=& \hat{B}_{\ell}^{\mu}(x_{\trans}^{},iz\beta) ~,
\end{eqnarray}
where we assumed, that at $\tau_{0}^{}$ Schr\"odinger and Heisenberg
operators coincide. The free evolution operator $U_{0}^{}$ is
given by
\begin{eqnarray}
\label{FreeEvOp}
U_{0}^{}(\tau,\tau_{0}^{};n)
= {\rm e}^{-i \hat{H}_{s}^{}(n) (\tau - \tau_{0}^{})} ~.
\end{eqnarray}
Making use of Eq.~(\ref{B:Heisenberg}) in Eq.~(\ref{Rho:RelLin}) we
find the relevant statistical operator
\begin{eqnarray}
\label{Rho:RelLin1}
\varrho_{{\rm rel}}^{}(n,\tau) 
= \Bigg( 1 + \beta\sum_{\ell} \int_{\sigma_{n}^{}} d\sigma \;
\phi_{\mu}^{\ell}(x)
\int_{0}^{1} dz \, \hat{B}_{\ell}^{\mu}(x_{\trans}^{},iz\beta)
\Bigg)
\varrho_{0}^{}(n)
\end{eqnarray}
and finally Eq.~(\ref{Rho:RelLin1F}) is obtained after Fourier
transformation. 

\subsection*{The irrelevant statistical operator in linear response}
In order to express $\Delta\varrho (n,\tau)$ [see
Eq.~(\ref{DeltaRho})] in linear response we first consider the commutator
\begin{equation}
\label{C:RhoRel,H_1}
\left[ {\varrho}^{}_{\rm rel}(n,\tau),
\hat{H}^{\tau}(n) \right]
= \left[ {\varrho}^{}_{\rm rel}(n,\tau),
\hat{H}_{s}^{}(n) + \hat{H}_{{\rm ext}}^{\tau}(n) \right]
\end{equation}
and note, that the zero order part
\begin{math}
\left[ {\varrho}^{}_{0}(n),
\hat{H}_{s}^{}(n) \right] = 0
\end{math}
vanishes by definition. There are two first order terms
\begin{eqnarray}
\label{C:RhoRel,H_2}
\left[ {\varrho}^{}_{0}(n), 
\hat{H}_{{\rm ext}}^{\tau}(n) \right]
\qquad\mbox{and}\qquad
\left[ {\varrho}^{(1)}_{\rm rel}(n,\tau),
\hat{H}_{s}^{}(n) \right] ~,
\end{eqnarray}
where $\varrho^{(1)}_{\rm rel}$ is the first order contribution of 
$\varrho^{}_{\rm rel}$.
Making use of the Kubo-identity
\begin{eqnarray}
\label{Kubo:Ident}
\left[ \hat{C}_{2}^{} , {\rm e}_{}^{\hat{C}_{1}^{}} \right]
= \int_{0}^{1} dz\, {\rm e}_{}^{z\hat{C}_{1}^{}}
\left[\hat{C}_{2}^{} , \hat{C}_{1}^{} \right]
{\rm e}_{}^{-z\hat{C}_{1}^{}} {\rm e}_{}^{\hat{C}_{1}^{}} ~,
\end{eqnarray}
we find
\begin{eqnarray}
\label{JHextComm1}
\left[ \varrho_{0}^{}(n),\hat{H}_{{\rm ext}}^{\tau}(n) \right]
&=& \beta \int_{\sigma_{n}^{}} d\sigma \, A_{{\rm ext}}^{\mu}(x)
\int_{0}^{1} dz \,
{\rm e}_{}^{-z \beta \left( \hat{H}_{s}^{} 
- \nu\hat{Q}_{}^{} \right)}
\nonumber\\[4pt]
& &
\times \left[ \,\hat{\!j}_{\mu}^{} , \left( \hat{H}_{s}^{} 
- \nu\hat{Q}_{}^{} \right) \right]
{\rm e}_{}^{z\beta \left( \hat{H}_{s}^{} 
- \nu\hat{Q}_{}^{} \right)} \varrho_{0}^{}(n) ~.
\end{eqnarray} 
Since
\begin{math}
\left[\,\hat{\!j}_{\mu}^{} , \hat{Q}_{}^{} \right] 
= n_{}^{\alpha}\left[\,\hat{\!j}_{\mu}^{} , \,\hat{\!j}_{\alpha}^{} \right] 
= 0 
\end{math}
and also
\begin{math}
\left[\hat{H}_{s}^{} , \hat{Q}_{}^{} \right] = 0 
\end{math}
($\hat{Q}$ is an integral of motion)
we can write
\begin{eqnarray}
\label{JHextComm2}
\left[ \varrho_{0}^{}(n),\hat{H}_{{\rm ext}}^{\tau}(n) \right]
= \beta \int_{\sigma_{n}^{}} d\sigma \, A_{{\rm ext}}^{\mu}(x)
\int_{0}^{1} dz \,
{\rm e}_{}^{-z \beta \hat{H}_{s}^{}}
\left[ \,\hat{\!j}_{\mu}^{} , \hat{H}_{s}^{} \right]
{\rm e}_{}^{z\beta \hat{H}_{s}^{}} \varrho_{0}^{}(n) ~.
\end{eqnarray}
Since the external field $A_{{\rm ext}}^{\mu}(x)$ appears in
(\ref{JHextComm2}) the commutator 
$\left[ \,\hat{\!j}_{\mu}^{} , \hat{H}_{s}^{} \right]$
is only treated in zero order, and we can write
\begin{eqnarray}
\label{HBerg1}
\partial_{\tau}^{} \,\hat{\!j}_{\mu}^{} 
= -i\left[ \,\hat{\!j}_{\mu}^{} ,
\hat{H} \right]
= -i\left[ \,\hat{\!j}_{\mu}^{} ,
\hat{H}_{s}^{} \right] ~.
\end{eqnarray}
Making use of Eq.~(\ref{B:Heisenberg}), we find
\begin{eqnarray}
\label{C:Rho0,HExt_3}
\left[ \varrho_{0}^{}(n),\hat{H}_{{\rm ext}}^{\tau}(n) \right]
&=& i\beta \int_{\sigma_{n}^{}} d\sigma \, A_{{\rm ext}}^{\mu}(x)
\int_{0}^{1} dz
\,\dot{\hat{\!j}}_{\!\mu}^{}(x_{\!\trans}^{},i\beta z)
\varrho_{0}^{}(n) ~,
\\[4pt]
\label{C:Rho0,HExt_4}
&=& i\beta {\rm e}_{}^{-ik_{\longi}^{}\tau} \int_{0}^{1} dz\,
A_{{\rm ext}}^{\mu}(k) \,
\dot{\hat{\!j}}_{\!\mu}^{\dagger} (k_{\!\trans}^{},i\beta z)
\varrho_{0}^{}(n) ~,
\end{eqnarray} 
with the last expression written in Fourier space.

The second commutator in Eq.~(\ref{C:RhoRel,H_2}) can be written 
in linear response by making use of Eq.~(\ref{Kubo:Ident})
\begin{eqnarray}
\label{C:RhoRel,H0_1}
\left[ {\varrho}^{(1)}_{\rm rel}(n,\tau),
\hat{H}_{s}^{}(n) \right] 
&=& i\beta \sum_{\ell} \int_{\sigma_{n}^{}} d\sigma \;
\phi_{\mu}^{\ell}(x)
\int_{0}^{1} dz 
\!\!\dot{\,\,\hat{B}_{\ell}^{\mu}}(x_{\trans}^{},iz\beta)
\varrho_{0}^{}(n) ~.
\end{eqnarray}
In Fourier representation Eq.~(\ref{C:RhoRel,H0_1}) reads
\begin{eqnarray}
\label{C:RhoRel,H0_2}
\left[ {\varrho}^{(1)}_{\rm rel}(n,\tau),
\hat{H}_{s}^{}(n) \right] 
&=& i\beta \sum_{\ell} {}
\phi_{\mu}^{\ell}(k)
\int_{0}^{1} dz 
\!\!\dot{\,\,\,\hat{B}_{\ell}^{\dagger\mu}}(k_{\trans}^{},iz\beta)
\varrho_{0}^{}(n) ~.
\end{eqnarray}

The derivative term in Eq.~(\ref{DeltaRho}) is calculated using 
Eq.~(\ref{Rho:RelLin1F})
\begin{eqnarray}
\label{DTau:RhoRel_1}
\partial_{\tau}^{} \varrho_{{\rm rel}}^{}(n,\tau) 
&=& -i\beta k_{\longi}^{} {\rm e}_{}^{-i k_{\longi}^{}\cdot \tau} 
\int_{0}^{1} dz \,
\sum_{\ell} 
\phi_{\mu}^{\ell}(k)\, \hat{B}_{\ell}^{\dagger\mu}(k_{\trans}^{},iz\beta)\, 
\varrho_{0}^{}(n) ~.
\end{eqnarray}

Constructing the expression for $\Delta\varrho_{}^{}(n,\tau)$ in the linear 
approximation [Eq.~(\ref{DeltaRho})], the evolution operators
$U$ are to be taken in zeroth order and yield a
$\tau$-translation. 
From the Eqns.~(\ref{DeltaRho}), (\ref{C:Rho0,HExt_4}),
(\ref{C:RhoRel,H0_1}), (\ref{DTau:RhoRel_1}) one finds
\begin{eqnarray}
\label{Rho:IrrLin1}
\Delta\varrho_{}^{}(n,\tau)
&=& -\beta \int_{-\infty}^{\tau} d\tau' {\rm e}_{}^{-\eta (\tau - \tau')}
{\rm e}_{}^{-ik_{\longi}^{}\tau'}
\int_{0}^{1} dz
\nonumber \\[2mm]
& &
\bigg\{ \sum_{\ell} \Big[
\!\!\!\dot{\,\,\,\hat{B}_{\ell}^{\dagger\mu}}(\tau'-\tau+iz\beta)
-ik_{\longi}^{} \hat{B}_{\ell}^{\dagger\mu}(\tau'-\tau+iz\beta)
\Big] \phi_{\mu}^{\ell}(k)
\nonumber \\[2mm]
& &
+ A_{{\rm ext}}^{\mu}(k) 
\,\dot{\hat{\!j}}_{\!\mu}^{} (\tau' - \tau + iz\beta)
\bigg\} \varrho_{0}^{}(n) ~.
\end{eqnarray}
Eq.~(\ref{Rho:IrrLin1}) can be rewritten using the transformation
$\tilde{\tau} = \tau - \tau'$ and Eq.~(\ref{Rho:IrrLin2F}) is obtained.

\renewcommand{\theequation}{C.\arabic{equation}}
\setcounter{equation}{0}
\section*{Appendix C}
\subsection*{The RPA susceptibility tensor}
We evaluate the current-force correlation function in the RPA
approximation with Eq.~(\ref{TotElCurr}), (\ref{JDotDir3}) as well as
\begin{math}
\,\dot{\hat{\!j}}_{\!\nu}^{\dagger}(\vek{k})
= \,\dot{\hat{\!j}}_{\!\nu}^{}(-\vek{k})
\end{math}
and the short-hand notation
\begin{math}
\langle \,\hat{\!j}_{\!\mu}^{}(\vek{k})\, ;
\,\dot{\hat{\!j}}_{\!\nu}^{}(\vek{k}) \rangle_{\omega+i\eta}^{}
\equiv \langle \,\hat{\!j}_{\!\mu}^{}\, ;
\,\dot{\hat{\!j}}_{\!\nu}^{} \rangle_{\omega+i\eta}^{\vekind{k}}
\end{math}
\begin{eqnarray}
\label{JJDot_Dir1}
\lefteqn{
\Big\langle \,\hat{\!j}_{\!\mu}^{}\, ;
\,\dot{\hat{\!j}}_{\!\nu}^{} \Big\rangle_{\omega+i\eta}^{\vekind{k}}
\equiv \frac{1}{\beta}\int_{0}^{\infty} d\bar{t}\;
{\rm e}_{}^{i(\omega+i\eta)\bar{t}}
\int_{0}^{\beta} d \tilde{t}\; {\rm Tr}\Big\{ 
\,\hat{\!j}_{\!\mu}^{}(\vek{k},\bar{t}-i\hbar\tilde{t}) 
\,\dot{\hat{\!j}}_{\!\nu}^{}(-\vek{k}) \varrho_{0}^{}\Big\} }
\nonumber\\[1pt]
&=&
-\frac{ie^2m^2}{\hbar\beta} \int_{0}^{\infty} d\bar{t}\;
{\rm e}_{}^{i(\omega+i\eta)\bar{t}}
\int_{0}^{\beta} d \tilde{t}
\int d_{}^{3}p\; d_{}^{3}p' \;
\sqrt{\frac{1}{E_{p}^{}}}
\sqrt{\frac{1}{E_{p'}^{}}} \,
\sum_{ss'rr'} \Bigg\{
\nonumber\\[1pt]
&(-1)&                                         
\sqrt{\frac{1}{E_{p+\hbar k}^{}}}
\sqrt{\frac{1}{E_{p'-\hbar k}^{}}}
{\rm Tr}\Big\{ 
\spar{-33}{78}{\spar{-35}{40}{ \Norder
\hat{d}_{p,s}^{\dagger}\hat{d}_{p+\hbar k,s'}^{}\Norder\,}}%
{{\Norder \hat{d}_{p',r}^{\dagger} \hat{d}_{p'-\hbar k,r'}^{}\Norder\,}}
\varrho_{0}^{} \Big\}\;
\nonumber\\
& &
{\rm tr}_{D}\Big\{ \Big( \bar{v}(p+\hbar k,s') \gamma_{\mu}^{} v(p,s) \Big) \,
\Big( \bar{v}(p'-\hbar k),r') \gamma_{\nu}^{} v(p',r) \Big) \Big\}
\nonumber\\
& &
\Big[E_{p'}^{}-E_{p'-\hbar k}^{}\Big]\,
{\rm e}_{}^{\frac{i}{\hbar}(E_{p}^{}-E_{p+\hbar k}^{})(\bar{t}-i\hbar\tilde{t})}
\nonumber\\[1pt]
&+&                                            
\sqrt{\frac{1}{E_{p}^{}-\hbar k}}
\sqrt{\frac{1}{E_{p'}^{}+\hbar k}} \,
{\rm Tr}\Big\{ 
\spar{-17}{95}{\spar{-21}{20}{ \Norder 
\hat{b}_{p-\hbar k,s'}^{\dagger} \hat{b}_{p,s}^{}\Norder\,}}%
{{\Norder \hat{b}_{p'+\hbar k,r'}^{\dagger} \hat{b}_{p',r}^{}\Norder \,}}
\varrho_{0}^{} \Big\}\;
\nonumber\\
& & 
{\rm tr}_{D}\Big\{ \Big( \bar{u}(p-\hbar k,s') \gamma_{\mu}^{} u(p,s) \Big) \,
\Big( \bar{u}(p'+\hbar k,r') \gamma_{\nu}^{} u(p',r) \Big) \Big\}
\nonumber\\
& & 
\Big[E_{p'}^{}-E_{p'+\hbar k}^{}\Big]\,
{\rm e}_{}^{-\frac{i}{\hbar}(E_{p}^{}-E_{p-\hbar k}^{})
(\bar{t}-i\hbar\tilde{t})}
\nonumber\\[1pt]
&+&                                             
\sqrt{\frac{1}{E_{p-\hbar k}^{}}}
\sqrt{\frac{1}{E_{p'-\hbar k}^{}}}
{\rm Tr}\Big\{ 
\spar{-60}{63}{\spar{-2}{40}{\Norder 
\hat{d}_{-p+\hbar k,s'}^{} \hat{b}_{p,s}^{}\Norder \,}}%
{{\Norder \hat{d}_{p',r}^{\dagger} 
\hat{b}_{-p'+\hbar k,r'}^{\dagger}\Norder\,}}
\varrho_{0}^{} \Big\}\;
\nonumber\\
& & 
{\rm tr}_{D}\Big\{ \Big( \bar{v}(-p+\hbar k,s') \gamma_{\mu}^{} u(p,s) \Big) \,
\Big( \bar{u}(-p'+\hbar k,r') \gamma_{\nu}^{} v(p',r) \Big) \Big\}
\nonumber\\
& & 
\Big[ E_{p'}^{} + E_{p'+\hbar k}^{}\Big]\,
{\rm e}_{}^{-\frac{i}{\hbar}(E_{p}^{}-E_{p-\hbar k}^{})
(\bar{t}-i\hbar\tilde{t})}
\nonumber\\[1pt]
&+&                                             
\sqrt{\frac{1}{E_{p+\hbar k}^{}}}
\sqrt{\frac{1}{E_{p'+\hbar k}^{}}}
{\rm Tr}\Big\{ 
\spar{-17}{110}{\spar{-19}{21}{\Norder 
\hat{b}_{-p-\hbar k,s'}^{\dagger}
\hat{d}_{p,s}^{\dagger}\Norder \,}}%
{{\Norder \hat{d}_{-p'-\hbar k,r'}^{} \hat{b}_{p',r}^{}\Norder \,}}
\varrho_{0}^{} \Big\}\;
\nonumber\\
& & 
{\rm tr}_{D}\Big\{ \Big( \bar{u}(-p-\hbar k,s') \gamma_{\mu}^{} v(p,s) \Big) \,
\Big( \bar{v}(-p'-\hbar k,r') \gamma_{\nu}^{} v(p',r) \Big)
\Big\} 
\nonumber\\
& & 
\Big[ E_{p'}^{} + E_{p'+\hbar k}^{}\Big]\,
{\rm e}_{}^{\frac{i}{\hbar}(E_{p}^{}+E_{p+\hbar k}^{})
(\bar{t}-i\hbar\tilde{t})}
\Bigg\} ~.
\end{eqnarray}
The notation ${\rm tr}_{D}^{}\{\ldots\}$ stands for the Dirac trace.
Making use of Wick's theorem~\cite{Gross,Evans}, the non-vanishing
contractions, which are denoted by the brackets 
$\spar{-13}{9}{\hat{b}_{ps}^{}}{\hat{b}_{p's'}^{\dagger}}$, can be
written in the form 
\begin{eqnarray}
\label{WTheorem}
& &
{\rm Tr}\Big\{
\hat{b}_{ps}^{\dagger},\hat{b}_{p's'}^{} \varrho_{0}^{} \Big\}
= \delta(\vek{p}-\vek{p}') \delta_{ss'}^{} f(E_{p}^{})
\nonumber\\[1pt]
& &
{\rm Tr}\Big\{
\hat{b}_{ps}^{},\hat{b}_{p's'}^{\dagger} \varrho_{0}^{} \Big\}
= \delta(\vek{p}-\vek{p}') \delta_{ss'}^{} \Big(1 - f(E_{p}^{}) \Big)
\nonumber\\[1pt]
& &
{\rm Tr}\Big\{
\hat{d}_{ps}^{\dagger},\hat{d}_{p's'}^{} \varrho_{0}^{} \Big\}
= \delta(\vek{p}-\vek{p}') \delta_{ss'}^{} \bar{f}(E_{p}^{}) 
\nonumber\\[1pt]
& &
{\rm Tr}\Big\{
\hat{d}_{ps}^{},\hat{d}_{p's'}^{\dagger} \varrho_{0}^{} \Big\}
= \delta(\vek{p}-\vek{p}') \delta_{ss'}^{} \Big(1 -
\bar{f}(E_{p}^{}) \Big) ~.
\end{eqnarray}
Further we introduced the Fermi-distribution functions $f$ and $\bar{f}$
for particles and antiparticles respectively 
\begin{eqnarray}
\label{FFunc}
f(E_{p}^{}) = \frac{1}{{\rm e}_{}^{\beta(E_{p}^{}-\mu)}+1} 
\qquad , \qquad
\bar{f}(E_{p}^{}) = \frac{1}{{\rm e}_{}^{\beta(E_{p}^{}+\mu)}+1} 
\end{eqnarray}
It should be noted that contractions inside the normal order do not
contribute and that crossing contraction lines (the third term in
Eq.~(\ref{JJDot_Dir1})) give an extra minus sign.
Eq.~(\ref{JJDot_Dir1}) can now be written in terms of Fermi functions
if the integration over $p'$ and the summation over $r$ and $r'$ is
performed
\begin{eqnarray}
\label{JJDot_Dir2}
\lefteqn{\Big\langle \,\hat{\!j}_{\!\mu}^{}\, ;
\,\dot{\hat{\!j}}_{\!\nu}^{} \Big\rangle_{\omega+i\eta}^{\vekind{k}}
=
-\frac{ie^2m^2}{\hbar\beta} \int_{0}^{\infty} d\bar{t}\;
{\rm e}_{}^{i(\omega+i\eta)\bar{t}}
\int_{0}^{\beta} d \tilde{t}
\int \frac{d_{}^{3}p}{(2\pi\hbar)_{}^{3}}\;
\frac{1}{E_{p}^{}}
\sum_{ss'} \Bigg\{ }
\nonumber\\[1pt]
&(-1)&                                           
\frac{1}{E_{p+\hbar k}^{}}\,
\bar{f}(E_{p}^{})\Big[ 1-\bar{f}(E_{p+\hbar k}^{}) \Big]\;
\Big[E_{p+\hbar k}^{}-E_{p}^{}\Big]\,
{\rm e}_{}^{\frac{i}{\hbar}(E_{p}^{}-E_{p+\hbar k}^{})
(\bar{t}-i\hbar\tilde{t})}
\nonumber\\
& & ~~
{\rm tr}_{D}\Big\{ \Big( \bar{v}(p+\hbar k,s') 
\gamma_{\mu}^{} v(p,s) \Big) \,
\Big( \bar{v}(p,s) \gamma_{\nu}^{} v(p+\hbar k,s' \Big) \Big\}
\nonumber\\[1pt]
&+&                                              
\frac{1}{E_{p-\hbar k}^{}}\,
f(E_{p-\hbar k}^{})\Big[ 1-f(E_{p}^{}) \Big]\;
\Big[E_{p-\hbar k}^{}-E_{p}^{}\Big]\,
{\rm e}_{}^{-\frac{i}{\hbar}(E_{p}^{}-E_{p-\hbar k}^{})
(\bar{t}-i\hbar\tilde{t})}
\nonumber\\
& & ~~
{\rm tr}_{D}\Big\{ \Big( 
\bar{u}(p-\hbar k,s') \gamma_{\mu}^{} u(p,s) \Big) \,
\Big( \bar{u}(p,s) \gamma_{\nu}^{} u(p-\hbar k,s') \Big) \Big\}
\nonumber\\[1pt]
&-&                                               
\frac{1}{E_{p-\hbar k}^{}}\,
\Big[ 1-\bar{f}(E_{p-\hbar k}^{}) \Big]\Big[ 1-f(E_{p}^{}) \Big] \;
\Big[ E_{p}^{} + E_{p-\hbar k}^{}\Big]\,
{\rm e}_{}^{-\frac{i}{\hbar}(E_{p}^{}+E_{p-\hbar k}^{})
(\bar{t}-i\hbar\tilde{t})}
\nonumber\\
& & ~~
{\rm tr}_{D}\Big\{ \Big( 
\bar{v}(-p+\hbar k,s') \gamma_{\mu}^{} u(p,s) \Big) \,
\Big( \bar{u}(p,s) \gamma_{\nu}^{} v(-p+\hbar k,s') \Big) \Big\}
\nonumber\\[1pt]
&+&                                               
\frac{1}{E_{p+\hbar k}^{}}\,
f(E_{p+\hbar k}^{}) \bar{f}(E_{p}^{})\;
\Big[ E_{p}^{} + E_{p+\hbar k}^{}\Big]\,
{\rm e}_{}^{\frac{i}{\hbar}(E_{p}^{}+E_{p+\hbar k}^{})
(\bar{t}-i\hbar\tilde{t})}
\nonumber\\
& & ~~
{\rm tr}_{D}\Big\{ \Big( 
\bar{u}(-p-\hbar k,s') \gamma_{\mu}^{} v(p,s) \Big) \,
\Big( \bar{v}(p,s) \gamma_{\nu}^{} u(-p-\hbar k,s') \Big)
\Big\} \Bigg\} \, .
\end{eqnarray}
Finally the spin summation rules 
\begin{eqnarray}
\label{SpinSum_u}
& &
\sum_{s}u(p,s) \otimes \bar{u}(p,s)
=\frac{1}{2m}(\psl+m)
\\[2pt]
\label{SpinSum_v}
& &
\sum_{s}v(p,s) \otimes \bar{v}(p,s)
=\frac{1}{2m}(\psl-m)
\end{eqnarray}
and the integrations over $\bar{t}$ and $\tilde{t}$ can be performed in
Eq.~(\ref{JJDot_Dir2}).
With the relations
\begin{eqnarray}
\label{FermiRel:1}
& &
\bar{f}(E_{p}^{})\Big( 1 - \bar{f}(E_{p+\hbar k}^{}) \Big)
\Big( {\rm e}_{}^{+\beta(E_{p}^{}-E_{p+\hbar k}^{})} - 1 \Big)
= - \Big( \bar{f}(E_{p}^{}) - \bar{f}(E_{p+\hbar k}^{}) \Big)
\nonumber\\[3pt]
& &
f(E_{p-\hbar k}^{}) \Big( 1-f(E_{p}^{}) \Big) 
\Big( {\rm e}_{}^{-\beta(E_{p}^{}-E_{p-\hbar k}^{})} - 1 \Big)
= f(E_{p}^{}) - f(E_{p-\hbar k}^{})
\nonumber\\[3pt]
& &
\Big( 1-f(E_{p}^{}) \Big) \Big( 1-\bar{f}(E_{p-\hbar k}^{}) \Big)
\Big( {\rm e}_{}^{-\beta(E_{p}^{}-E_{p-\hbar k}^{})} - 1 \Big)
= -1 + f(E_{p}^{}) + \bar{f}(E_{p-\hbar k}^{})
\nonumber\\[3pt]
& &
\bar{f}(E_{p}^{}) f(E_{p+\hbar k}^{})
\Big( {\rm e}_{}^{+\beta(E_{p}^{}-E_{p+\hbar k}^{})} - 1 \Big)
= 1 - \bar{f}(E_{p}^{}) - f(E_{p+\hbar k}^{})
\end{eqnarray}
the equation~(\ref{Chi_RPA1}) is obtained. 

\renewcommand{\theequation}{D.\arabic{equation}}
\setcounter{equation}{0}
\section*{Appendix D}
\subsection*{Calculation of
$\,\dot{\hat{\!j}}_{\!\mu}^{}(\vek{k})$}
From the Eqs.~(\ref{TotElCurr}) and (\ref{HIon}) we find
\begin{eqnarray}
\label{JDotRWA_Ion1}
& & \,\,\dot{\hat{\!\!j}}_{\mu}^{}(\vek{-k})
= -\frac{ie}{\hbar} \int \frac{d_{}^{3}q}{(2\pi\hbar)_{}^{3}} \,
d_{}^{3}p \,
d_{}^{3}p_1 \,
\sqrt{\frac{m}{E_{p}^{}}} \,
\sqrt{\frac{m}{E_{p_{1}^{}}^{}}} \,
A_{ion}^{0}(\vek{q})
\sum_{ss'rr'} \bigg\{
\nonumber\\[2pt]
& & ~~                                          
\sqrt{\frac{m}{E_{p-\hbar k}}} \,
\sqrt{\frac{m}{E_{p_1^{}-q}}} \;
\Big[\Norder 
\hat{d}_{p-\hbar k,s'}^{} \hat{d}_{p,s}^{\dagger}\Norder\, ,
\Norder 
\hat{d}_{p_{1}^{}-q,r'}^{} \hat{d}_{p_{1}^{},r}^{\dagger} \Norder\Big] \;
{\rm e}_{}^{\frac{i}{\hbar}
(E_{p}^{}-E_{p-\hbar k}^{}+E_{p_{1}^{}}^{}-E_{p_{1}^{}-q}^{})t}
\nonumber\\[2pt]
& & \quad \times
\Big( \bar{v}(p+\hbar k,s') \gamma_{\mu}^{} v(p,s) \Big)
\Big( \bar{v}(p_{1}^{}-q,r') \gamma_{0}^{} v(p_{1}^{},r) \Big)\,
\nonumber\\[1pt]
& & +                                           
\sqrt{\frac{m}{E_{p+\hbar k}}} \,
\sqrt{\frac{m}{E_{p_1^{}+q}}} \;
\Big[\Norder 
\hat{b}_{p+\hbar k,s'}^{\dagger} \hat{b}_{p,s}^{}\Norder\, ,
\Norder 
\hat{b}_{p_{1}^{}+q,r'}^{\dagger} \hat{b}_{p_{1}^{},r}^{} \Norder\Big]\;
{\rm e}_{}^{\frac{i}{\hbar} 
(-E_{p}^{}+E_{p+\hbar k}^{}-E_{p_{1}^{}}^{}+E_{p_{1}^{}+q}^{})t}
\nonumber\\[1pt]
& & \quad \times
\Big( \bar{u}(p+\hbar k,s') \gamma_{\mu}^{} u(p,s) \Big)
\Big( \bar{u}(p_{1}^{}+q,r') \gamma_{0}^{} u(p_{1}^{},r) \Big)\,
\nonumber\\[1pt]
& & +                                           
\sqrt{\frac{m}{E_{p+\hbar k}}} \,
\sqrt{\frac{m}{E_{p_1^{}-q}}} \;
\Big[\Norder 
\hat{d}_{-p-\hbar k,s'}^{} \hat{b}_{p,s}^{}\Norder\, ,
\Norder 
\hat{d}_{p_{1}^{}-q,r'}^{} \hat{d}_{p_{1}^{},r}^{\dagger} \Norder\Big]\,
{\rm e}_{}^{\frac{i}{\hbar}
(-E_{p}^{}-E_{p+\hbar k}^{}+E_{p_{1}^{}}^{}-E_{p_{1}^{}-q}^{})t}
\nonumber\\[1pt]
& & \quad \times
\Big( \bar{v}(-p-\hbar k,s') \gamma_{\mu}^{} u(p,s) \Big)
\Big( \bar{v}(p_{1}^{}-q,r') \gamma_{0}^{} v(p_{1}^{},r) \Big)\,
\nonumber\\[1pt]
& & +                                           
\sqrt{\frac{m}{E_{p+\hbar k}}} \,
\sqrt{\frac{m}{E_{p_1^{}+q}}} \;
\Big[\Norder 
\hat{d}_{-p-\hbar k,s'}^{} \hat{b}_{p,s}^{}\Norder\, ,
\Norder 
\hat{b}_{p_{1}^{}+q,r'}^{\dagger} \hat{b}_{p_{1}^{},r}^{} \Norder\Big]\,
{\rm e}_{}^{\frac{i}{\hbar}
(-E_{p}^{}-E_{p+\hbar k}^{}-E_{p_{1}^{}}^{}+E_{p_{1}^{}+q}^{})t}
\nonumber\\[1pt]
& & \quad \times
\Big( \bar{v}(-p-\hbar k,s') \gamma_{\mu}^{} u(p,s) \Big)
\Big( \bar{u}(p_{1}^{}+q,r') \gamma_{0}^{} u(p_{1}^{},r) \Big)\,
\nonumber\\[1pt]
& & +                                           
\sqrt{\frac{m}{E_{p-\hbar k}}} \,
\sqrt{\frac{m}{E_{p_1^{}-q}}} \;
\Big[\Norder 
\hat{b}_{-p+\hbar k,s'}^{\dagger} \hat{d}_{p,s}^{\dagger}\Norder\, ,
\Norder 
\hat{d}_{p_{1}^{}-q,r'}^{} \hat{d}_{p_{1}^{},r}^{\dagger} \Norder\Big]\,
{\rm e}_{}^{\frac{i}{\hbar}
(E_{p}^{}+E_{p-\hbar k}^{}+E_{p_{1}^{}}^{}-E_{p_{1}^{}-q}^{})t}
\nonumber\\[1pt]
& & \quad \times
\Big( \bar{u}(-p+\hbar k,s') \gamma_{\mu}^{} v(p,s) \Big)
\Big( \bar{v}(p_{1}^{}-q,r') \gamma_{0}^{} v(p_{1}^{},r) \Big)\,
\nonumber\\[1pt]
& & +                                           
\sqrt{\frac{m}{E_{p-\hbar k}}} \,
\sqrt{\frac{m}{E_{p_1^{}+q}}} \;
\Big[\Norder 
\hat{b}_{-p+\hbar k,s'}^{\dagger} \hat{d}_{p,s}^{\dagger}\Norder\, ,
\Norder 
\hat{b}_{p_{1}^{}+q,r'}^{\dagger} \hat{b}_{p_{1}^{},r}^{} \Norder\Big]\,
{\rm e}_{}^{\frac{i}{\hbar}
(E_{p}^{}+E_{p-\hbar k}^{}-E_{p_{1}^{}}^{}+E_{p_{1}^{}+q}^{})t}
\nonumber\\[1pt]
& & \quad \times
\Big( \bar{u}(-p+\hbar k,s') \gamma_{\mu}^{} v(p,s) \Big)
\Big( \bar{u}(p_{1}^{}+q,r') \gamma_{0}^{} u(p_{1}^{},r) \Big)\,
\bigg\} ~.
\end{eqnarray}
After calculating the commutators in Eq.~(\ref{JDotRWA_Ion1}), the
integration over $p_{1}^{}$ as well the summation over the primed
variable can be performed. Finally, using the spin summation
Eq.~(\ref{SpinSum_u}) and (\ref{SpinSum_v}), we find the following
equation for the force operator (only the non-vanishing terms in the
RWA-approximation are kept)
\begin{eqnarray}
\label{JDotRWA_Ion3}
& & \,\,\dot{\hat{\!\!j}}_{\mu}^{}(-\vek{k})
= -\frac{iem_{}^{}}{2\hbar} \int \frac{d_{}^{3}q}{(2\pi\hbar)_{}^{3}} \,
d_{}^{3}p \,
A_{ion}^{0}(\vek{q})
\sum_{ss'} \bigg\{
\nonumber\\[4pt]
& & ~~                                          
\sqrt{\frac{1}{E_{p}^{}}} \,
\sqrt{\frac{1}{E_{p-q-\hbar k}}} \;
{\rm e}_{}^{\frac{i}{\hbar}
(E_{p}^{}-E_{p-q-\hbar k}^{})t} \,
\nonumber\\[1pt]
& & \quad \times
\Big( \bar{v}(p-q-\hbar k,s') \gamma_{0}^{}
\frac{\psl-\hbar \ksl - m}{E_{p-\hbar k}}
\gamma_{\mu}^{} v(p,s) \Big) ~
\hat{d}_{p,s}^{\dagger} \hat{d}_{p-q-\hbar k,s'}^{}
\nonumber\\[2pt]
& & -                                           
\sqrt{\frac{1}{E_{p+q}^{}}} \,
\sqrt{\frac{1}{E_{p-\hbar k}}} \;
{\rm e}_{}^{\frac{i}{\hbar}
(E_{p+q}^{}-E_{p-\hbar k}^{})t} \,
\nonumber\\[1pt]
& & \quad \times
\Big( \bar{v}(p-\hbar k,s') \gamma_{\mu}^{}
\frac{\psl - m}{E_{p}^{}}
\gamma_{0}^{} v(p+q,s) \Big) ~
\hat{d}_{p+q,s}^{\dagger} \hat{d}_{p-\hbar k,s'}^{} 
\nonumber\\[1pt]
& & +                                           
\sqrt{\frac{1}{E_{p-q}}} \,
\sqrt{\frac{1}{E_{p+\hbar k}}} \;
{\rm e}_{}^{\frac{i}{\hbar} 
(E_{p+\hbar k}^{}-E_{p-q}^{})t}\,
\nonumber\\[1pt]
& & \quad \times
\Big( \bar{u}(p+\hbar k,s') \gamma_{\mu}^{}
\frac{\psl + m}{E_{p}}
\gamma_{0}^{} u(p-q,s) \Big) ~
\hat{b}_{p+\hbar k,s'}^{\dagger} \hat{b}_{p-q,s}^{}
\nonumber\\[2pt]
& & -                                           
\sqrt{\frac{1}{E_{p}}} \,
\sqrt{\frac{1}{E_{p+q+\hbar k}}} \;
{\rm e}_{}^{-\frac{i}{\hbar} 
(E_{p}-E_{p+q+\hbar k}^{})t} \,
\nonumber\\[1pt]
& & \quad \times
\Big( \bar{u}(p+q+\hbar k,s') \gamma_{0}^{}
\frac{\psl+\hbar\ksl + m}{E_{p+\hbar k}}
\gamma_{\mu}^{} u(p,s) \Big) ~
\hat{b}_{p+q+\hbar k,s'}^{\dagger} \hat{b}_{p,s}^{}
\end{eqnarray}
We perform the shift $p \to p+\hbar k$ in the first term and 
$p \to p - \hbar k$ in the last term of Eq.~(\ref{JDotRWA_Ion3})
and introduce the initial and final momentum variables, $p_{i}^{}$
and $p_{f}^{}$ according to
\begin{eqnarray}
\label{DefPiPf}
q = p_{f}^{} - p_{i}^{} - \hbar k
\qquad , \qquad
\begin{array}{lcl}
p_{i}^{} = p - \hbar k
&\quad , \quad&
p_{f}^{} = p + q
\\
p_{i}^{} = p - q
&\quad , \quad&
p_{f}^{} = p+ \hbar k
\end{array}
~.
\end{eqnarray}
Using these redefinitions we can obtain Eqs.~(\ref{JDotRWA_Ion4}) --
(\ref{DefN2}) in terms of the transition matrices $N_{\mu}^{(e)}$ and
$N_{\mu}^{(p)}$ for electrons and positrons
respectively.


\section{Acknowledgments}
The main part of this work was conducted during visits in Rostock and
Moscow. V.G.~Morozov would like to thank the ``Deutsche
Forschungsgemeinschaft'' and A.~H\"oll the ``Studienstiftung des deutschen 
Volkes'' and the ``Deutsche Forschungsgemeinschaft'' for supporting
this work.

\end{document}